\documentclass[12pt,preprint]{emulateapj}

\def\dndz{({\rm d}\log_{10}\dot{N}_{{\rm ion}}/dz)_{z=8}}

\begin{document}
\title{Reionization after Planck: The Derived Growth of the Cosmic
  Ionizing Emissivity now matches the Growth of the Galaxy UV
  Luminosity Density}
\author{R. J. Bouwens\altaffilmark{2,3},
 G. D. Illingworth\altaffilmark{3}, 
 P. A. Oesch\altaffilmark{4},
 J. Caruana\altaffilmark{5},
 B. Holwerda\altaffilmark{2},
 R. Smit\altaffilmark{6},
 S. Wilkins\altaffilmark{7}}

\altaffiltext{1}{Based on observations made with the NASA/ESA Hubble
  Space Telescope, which is operated by the Association of
  Universities for Research in Astronomy, Inc., under NASA contract
  NAS 5-26555.}  \altaffiltext{2}{Leiden Observatory, Leiden
  University, NL-2300 RA Leiden, Netherlands}
\altaffiltext{3}{UCO/Lick Observatory, University of California, Santa
  Cruz, CA 95064}
\altaffiltext{4}{Department of Astronomy, Yale University, New Haven, CT 06520}
\altaffiltext{5}{Leibniz-Institut fur Astrophysik Potsdam (AIP), 
An der Sternwarte 16, D-14482 Potsdam, Germany}
\altaffiltext{6}{Department of Physics and Astronomy, South Road, Durham, DH1 3EE, United Kingdom}
\altaffiltext{7}{Department of Physics \& Astronomy, University of Sussex, Falmer, BRIGHTON, BN1 9QH, United Kingdom}

\begin{abstract}
Thomson optical depth $\tau$ measurements from Planck provide new
insights into the reionization of the universe.  In pursuit of
model-independent constraints on the properties of the ionizing
sources, we determine the empirical evolution of the cosmic ionizing
emissivity.  We use a simple two-parameter model to map out the
evolution in the emissivity at $z\gtrsim$6 from the new Planck optical
depth $\tau$ measurements, from the constraints provided by quasar
absorption spectra and from the prevalence of Lyman $\alpha$ emission
in $z\sim7$-8 galaxies.  We find the redshift evolution in the
emissivity $\dot{N}_{\rm ion}(z)$ required by the observations to be
$\dndz=-0.15_{-0.11}^{+0.08}$ ($\dndz=-0.19_{-0.11}^{+0.09}$ for a
flat prior), largely independent of the assumed clumping factor
$C_{HII}$ and entirely independent of the nature of the ionizing
sources.  The trend in $\dot{N}_{\rm ion}(z)$ is well-matched by the
evolution of the galaxy $UV$-luminosity density
(d$\log_{10}\rho_{UV}/dz=-0.11\pm0.04$) to a magnitude limit
$\gtrsim-$13 mag, suggesting that galaxies are the sources that drive
the reionization of the universe.  The role of galaxies is further
strengthened by the conversion from the UV luminosity density
$\rho_{UV}$ to $\dot{N}_{\rm ion}(z)$ being possible for
physically-plausible values of the escape fraction $f_{esc}$, the
Lyman-continuum photon production efficiency $\xi_{\rm ion}$, and
faint-end cut-off $M_{lim}$ to the luminosity function.  Quasars/AGN
appear to match neither the redshift evolution nor normalization of
the ionizing emissivity.  Based on the inferred evolution in the
ionizing emissivity, we estimate that the $z\sim10$ UV-luminosity
density is $8_{-4}^{+15}\times$ lower than at $z\sim6$, consistent
with the observations.  The present approach of contrasting the
inferred evolution of the ionizing emissivity with that of the galaxy
UV luminosity density adds to the growing observational evidence that
faint, star-forming galaxies drive the reionization of the universe.
\end{abstract}

\section{Introduction}

One of the most important phase transitions in the history of the
universe is the reionization of the neutral hydrogen gas.  Following
recombination early in the universe shortly after the Big Bang, the
universe likely remained in a largely neutral state until
$z\sim15$-25.  The collapse of the first dark matter halos and gas
cooling brought about the formation of the first stars and galaxies.
These early stars and galaxies have long been thought to provide the
ionizing photons necessary to reionize the universe (Loeb \& Barkana
2001; Loeb 2006).

Despite this general picture of reionization and the likely role that
early galaxies played in the process, establishing that this is the
case has been particularly challenging, both due to the difficulties
in probing the ionization state of the universe at $z>6$ (e.g., Ouchi
et al.\ 2010; Stark et al.\ 2010) and providing constraints on the
ionizing photons that early galaxies themselves are thought to produce
(e.g., Siana et al.\ 2010, 2015; Vanzella et al.\ 2012; Nestor et
al.\ 2013; Mostardi et al.\ 2013; Cooke et al.\ 2014).  Further
complicating the interpretation were early measurements of the Thomson
optical depth by WMAP (e.g., $\tau=0.17\pm0.06$: Spergel et al.\ 2003)
which suggested a significant quantity of the ionizing photons in the
early universe -- implying an instantaneous reionization redshift of
$z_r=20_{-9}^{+11}$ (95\% confidence) -- and which showed no clear
connection with the early evolution of galaxies.  While lower optical
depths were measured in subsequent studies by WMAP (i.e.,
$\tau=0.089\pm0.014$: Bennett et al.\ 2013) with an implied
$z_r=10.6\pm1.1$, these depths still pointed towards substantial
amounts of ionizing radiation being present in the early universe.

\begin{deluxetable*}{ccccc}
\tablewidth{0cm}
\tabletypesize{\footnotesize}
\tablecaption{Key Observational Constraints on the Reionization History of the Universe.\tablenotemark{a}\label{tab:obsconst}}
\tablehead{\colhead{Constraint} & \colhead{} & \colhead{} & \colhead{} & \colhead{}\\
\colhead{\#} & \colhead{Redshift} & \colhead{$Q_{HII}$ Constraint} & \colhead{Technique} & \colhead{Reference}}
\startdata
\\
\multicolumn{5}{c}{Key Constraints on the Ionization History of the Universe Explicitly Considered Here}\\
\multicolumn{5}{c}{In Deriving the Evolution of the Cosmic Ionizing Emissivity at $z>6$}\\\\
1. & & \multicolumn{2}{c}{Thomson Optical Depth $\tau = 0.066\pm0.013$} & PC15 \\
\\
2. & \multicolumn{4}{c}{Reionization Finishes Between $z=5.9$ and $z=6.5$}\\
& 5.03 & 0.9999451$_{-0.0000165}^{+0.0000142}$ & Gunn-Peterson Optical Depth & Fan et al.\ (2006a)\\
& 5.25 & 0.9999330$_{-0.0000244}^{+0.0000207}$ & Gunn-Peterson Optical Depth & Fan et al.\ (2006a)\\
& 5.45 & 0.9999333$_{-0.0000301}^{+0.0000247}$ & Gunn-Peterson Optical Depth & Fan et al.\ (2006a)\\
& 5.65 & 0.9999140$_{-0.0000460}^{+0.0000365}$ & Gunn-Peterson Optical Depth & Fan et al.\ (2006a)\\
& 5.85 & 0.9998800$_{-0.0000490}^{+0.0000408}$ & Gunn-Peterson Optical Depth & Fan et al.\ (2006a)\\
& 6.10 & 0.99957$\pm$0.00030 & Gunn-Peterson Optical Depth & Fan et al.\ (2006a)\\
& 5.9 & $>$0.89 & Dark Gaps in Quasar Spectra & McGreer et al.\ (2015)\\
& 5.6 & $>$0.91 & Dark Gaps in Quasar Spectra & McGreer et al.\ (2015)\\
& 6.24-6.42 & $<$0.9 (2$\sigma$) & Ly$\alpha$ Damping Wing of Quasars & Schroeder et al.\ (2013)\\
\\
\multicolumn{5}{c}{Higher-Redshift Constraints}\\
3. & 7.0 & $Q_{HII}(z=7) = 0.66_{-0.09}^{+0.12}$ & Prevalence of Ly$\alpha$ Emission in Galaxies & S14 \\
4. & 8.0 & $Q_{HII}(z=8)<0.35$ & Prevalence of Ly$\alpha$ Emission in Galaxies & S14 \\
\\
\multicolumn{5}{c}{Continuity with Ionizing Emissivity Estimates at $z=4.75$}\\
5. & \multicolumn{3}{c}{$\log_{10} \dot{N}_{\rm ion} (z=4.75) = 10^{50.99\pm0.45}$ s$^{-1}$ Mpc$^{-3}$} & BB13\\
\\
\multicolumn{5}{c}{Other Constraints on the Ionization History of the Universe Not Explicitly Used\tablenotemark{b}}\\
& 6.3 & $\geq$0.5 & Ly$\alpha$ Damping Wing of a GRB & Totani et al.\ (2006) \\
 & & & & McQuinn et al.\ (2008) \\
& 6.6 & $\geq$0.6 & Ly$\alpha$ Emitters & Ouchi et al.\ (2010) \\
& 6.6 & $\geq$0.5 & Galaxy Clustering & McQuinn et al.\ (2007),\\
&     &           &                    &  Ouchi et al.\ (2010) \\
& 7.0 & 0.32-0.64 & Ly$\alpha$-Emitter LFs & Ota et al.\ (2008)\\
& 7.0 & $\sim$0.5 & Prevalence of Ly$\alpha$ Emission in Galaxies & Caruana et al.\ (2014)\\
& 7.0 & 0.1-0.4 & Prevalence of Ly$\alpha$ Emission in Galaxies & Ono et al.\ (2012) \\
& 7.0 & $<$0.49 & Prevalence of Ly$\alpha$ Emission in Galaxies & P14 \\
& 7.0 & $<$0.5 & Prevalence of Ly$\alpha$ Emission in Galaxies & R13\tablenotemark{c} \\
& 7.0 & $<$0.5 & Clustering of Ly$\alpha$ Emitting Galaxies & Sobacchi \& Mesinger (2015)\\
& 7.1 & $\leq$0.9 & Near-Zone Quasar & Mortlock et al.\ (2011),\\
&     &          &                   & Bolton et al.\ (2011)\\
& 8.0 & $<$0.70 & Prevalence of Ly$\alpha$ Emission in Galaxies & Tilvi et al.\ (2014) 
\enddata
\tablenotetext{a}{This table is a compilation of the constraints
  presented in the original papers under References, but with valuable
  guidance by the results presented in Figures 5 and 3 from R13 and
  R15, respectively.}

\tablenotetext{b}{While not explicitly considered in deriving the
  evolution of the cosmic ionizing emissivity, almost all of these
  constraints are satisfied for the typical reionization histories
  derived in this study (see right panel of Figure~\ref{fig:reion}).}
\tablenotetext{c}{R13 estimate this constraint on $Q_{HII}$ based on
  the observational results from Fontana et al.\ (2010), Pentericci et
  al.\ (2011), Schenker et al.\ (2012), and Ono et al.\ (2012) and the
  simulation results from McQuinn et al.\ (2007), Mesinger \&
  Furlanetto (2008), and Dijkstra et al.\ (2011).}
\end{deluxetable*}

Fortunately, substantial progress has been made over the last ten
years to better understand cosmic reionization.  Much of the progress
has been observational, through the better study of bright quasars and
improved statistics on Ly$\alpha$ emission in normal star-forming
galaxies, to better probe the ionization state of the $z=6$-9 universe
(e.g., Mortlock et al.\ 2011; Ono et al.\ 2012; Pentericci et
al.\ 2014 [P14]; Schenker et al.\ 2014 [S14]; Caruana et al.\ 2014).
The greater depths of probes for $z=6$-10 galaxies provided greater
confidence that galaxies could provide the necessary reservoir of
photons to reionize the universe (e.g., Oesch et al.\ 2010; Bunker et
al.\ 2010; Bouwens et al.\ 2010, 2011; Ellis et al.\ 2013; McLure et
al.\ 2013; Oesch et al.\ 2013; Bouwens et al.\ 2015 [B15]).  However,
theoretical progress has been similarly substantial, due to
significantly improved estimates of the clumping factor (Bolton \&
Haehnelt 2007; Pawlik et al.\ 2009, 2015; Finlator et al.\ 2012; Shull
et al.\ 2012) and ever more sophisticated simulations tracking the
reionization of the universe and the propagation of Ly$\alpha$ out of
the galaxies and into the IGM (e.g., Mesinger et al.\ 2015; Choudhury
et al.\ 2015).  Lastly, measurements of the integrated column of
ionized material to the last-scattering surface from 3-year Planck
mission yield $\tau=0.066\pm0.013$ (Planck Collaboration et al. 2015,
XIII [PC15]), implying that $z_r=8.8_{-1.2}^{+1.3}$ and suggesting
that current surveys may be uncovering the sources that led to the
reionization of the universe.

\begin{figure}
\epsscale{1.15} \plotone{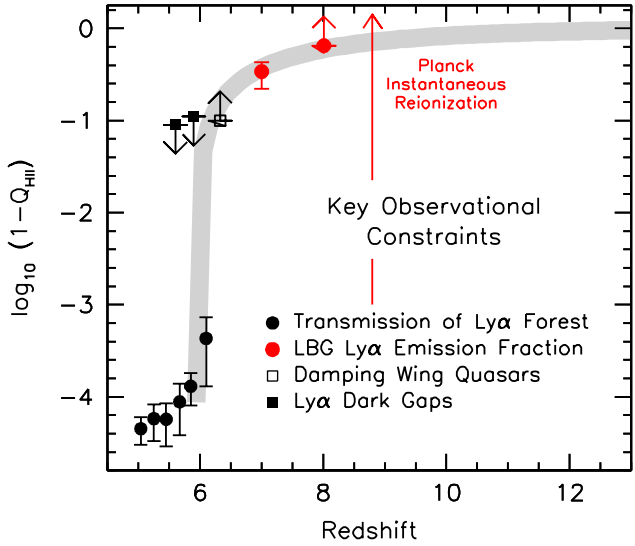}
\caption{The key observational constraints on the filling factor of
  ionized hydrogen $Q_{HII}$ considered here in modeling the evolution
  of the cosmic ionizing emissivity (\S2 and
  Table~\ref{tab:obsconst}).  These constraints include the
  Gunn-Peterson optical depths and dark-gap statistics measured in
  $z\sim6$ quasars (Fan et al.\ 2006a; McGreer et al.\ 2015), damping
  wings measured in $z\sim6.2$-6.4 quasars (Schroeder et al.\ 2013),
  and the prevalence of Ly$\alpha$ emission in $z\sim7$-8 galaxies
  (S14). Based on the constraints shown in this figure (particularly
  on the basis of the Gunn-Peterson troughs and dark-gap statistics and
  damping wings of $z\sim6$-6.5 quasars), reionization ends between
  $z=5.9$ and $z=6.5$.  The other key constraints we consider are the
  new Thomson optical depth measurements from Planck (PC15:
  represented here with a upward-pointing red arrow at the
  instantaneous reionization redshift $z_{reion}=8.8_{-1.2}^{+1.3}$)
  and a requirement for continuity with the cosmic ionizing emissivity
  at $z=4.75$, as derived by BB13.  The large changes in the filling
  factor are represented schematically by the grey-shaded
  region.\label{fig:obs}}
\end{figure}

Naturally, numerous studies have taken advantage of this collective
progress to construct self-consistent models for reionizing the
universe (e.g., Choudhury \& Ferrara 2006; Bolton \& Haehnelt 2007;
Oesch et al.\ 2009; Trenti et al.\ 2010; Haardt \& Madau et al.\ 2012
[HM12]; Bouwens et al.\ 2012a; Kuhlen \& Faucher-Gigu{\`e}re 2012
[KF12]; Shull et al.\ 2012; Finkelstein et al.\ 2012b; Alvarez et
al.\ 2012; Robertson et al.\ 2013 [R13]; Cai et al.\ 2014; Choudhury
et al.\ 2015; Ishigaki et al.\ 2015; Robertson et al.\ 2015 [R15]),
and it is indeed encouraging that many recent models (e.g., R15) prove successful in reionizing the universe, while
matching many other individual constraints on the reionization state
of the universe and also the consensus star formation history (e.g.,
Madau \& Dickinson 2014; B15).

While these analyses are reassuring and offer strong support for the
idea that galaxies reionize the universe, the \textit{uniqueness} of
galaxies as the source of photons to reionize the universe is more
challenging to establish.  Indeed, it is possible to imagine the
existence of other populations of ionizing sources (e.g., Chen et
al.\ 2003; Somerville et al.\ 2003; Hansen \& Haiman 2004; Madau et
al.\ 2004; Ricotti \& Ostriker 2004; Ricotti et al.\ 2008), which even
if speculative could also match current constraints.

What \textit{model-independent} statements can be made about the
sources that reionize the universe?  To answer this question, it is
useful to look at the \textit{evolution} of the cosmic ionizing
emissivity, since this allows us to keep assumptions regarding the
nature of the ionizers to a minimum.  As we will show, interesting
constraints on the evolution of the emissivity can be obtained based
on current observations, e.g., from the Thomson optical depths or the
inferred filling factor of ionized hydrogen $Q_{HII}$ at $z=6$-9 (see
also Mitra et al.\ 2011, 2012, 2013).

The purpose of the present analysis is to take advantage of current
observational constraints on the ionization state of the universe at
$z>6$ (e.g., Fan et al.\ 2006a; S14; PC15) to constrain the evolution
of the cosmic ionizing emissivity with redshift.  Through comparisons
with the evolution of the $UV\,$luminosity density of galaxies and
other potential ionizing sources, we can evaluate the likelihood that
each of these sources of ionizing photons drives the reionization of
the universe.  We begin with a description of the relevant
observations (\S2) and methodology (\S3) and then derive constraints
on the evolution of the ionizing emissivity $\dot{N}_{\rm ion}$
(\S4.1).  After deriving constraints on $\dot{N}_{\rm ion}(z)$, we
compare our results with what we would expect for galaxies (\S4.2),
quasars (\S4.3), and consider the implications for the $UV$ luminosity
density at $z\sim10$ (\S4.4).  We conclude with a brief summary (\S5).
We take $H_0=67.51\pm0.64$, $\Omega_{\Lambda}=0.6879\pm0.0087$, and
$\Omega_{m}=0.3121\pm0.0087$, $\Omega_{b}h^2=0.02230\pm0.00014$
(PC15).

\section{Observational Constraints}

There are a wide variety of observational constraints in the
literature on the ionization state of neutral hydrogen that can be
leveraged in considering questions regarding the reionization of the
universe.

\begin{figure*}
\epsscale{1.17}
\plotone{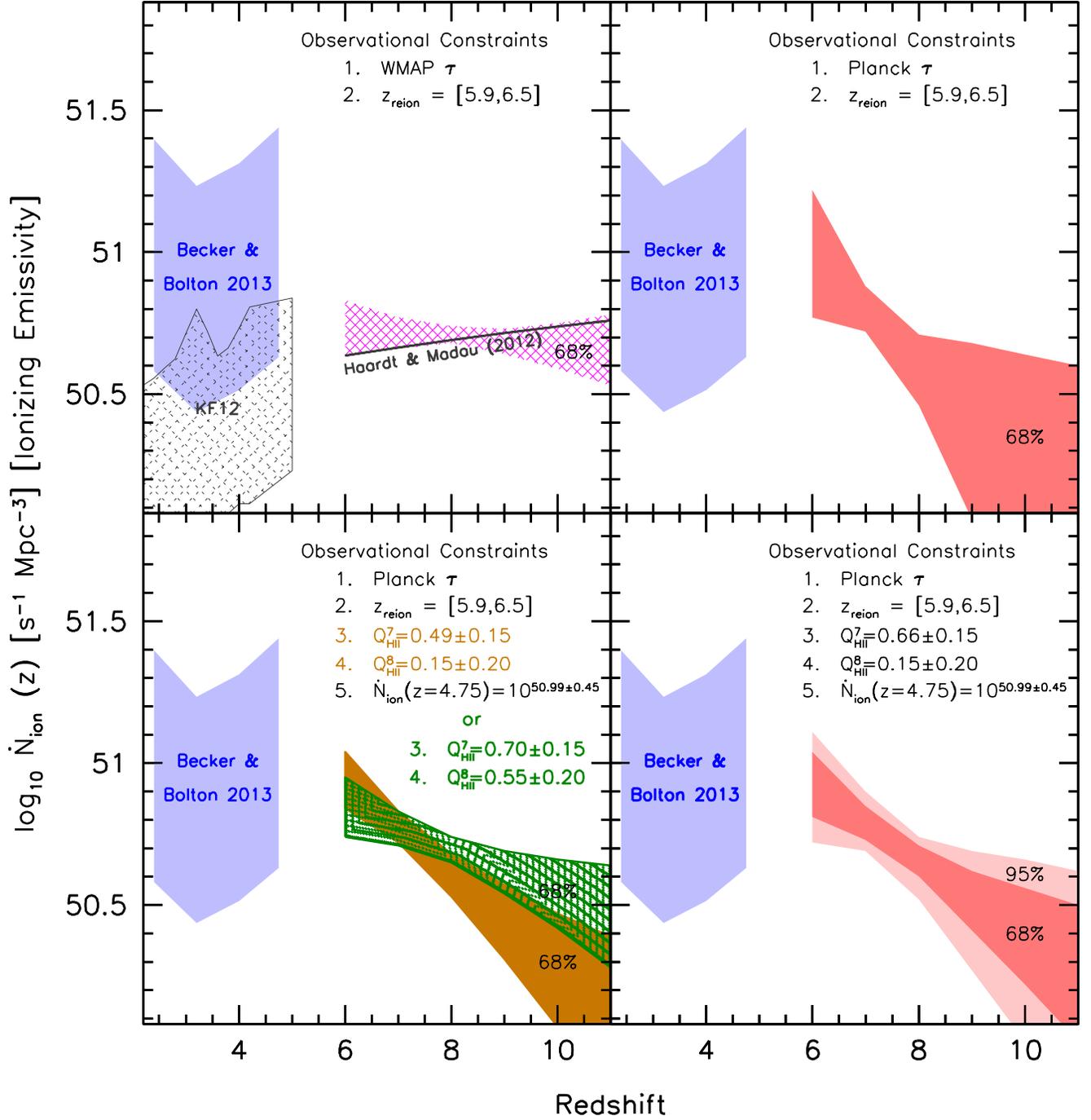}
\caption{Allowed evolution (68\% confidence) of the cosmic ionizing
  emissivity over the redshift range $z=6$ to $z=11$, as inferred from
  the observations.  (\textit{upper left}) Constraints on the
  evolution of the emissivity supposing a completion of reionization
  at $z=5.9$-6.5 (Fan et al.\ 2006a; Schroeder et al.\ 2013; McGreer
  et al.\ 2015) and the 9-year measured Thomson optical depth
  $\tau=0.089\pm0.14$ from WMAP (Bennett et al.\ 2013), and a simple
  modeling of the evolution of $Q_{HII}$ with $C_{HII}=3$.  The region
  so defined is presented in hatched magenta.  Constraints on the
  emissivity from $z=2$ to $z=5$ from BB13 are shown with the
  light-blue-shaded region and were derived from observations of the
  Ly$\alpha$ forest.  Systematic uncertainties in the inferred
  emissivity can often be large, $\sim0.3$ dex, due to the sensitivity
  to the assumed (or measured) temperature of the IGM and the opacity
  to ionizing photons.  As a result, some past constraints have been
  much lower than those from BB13 (e.g., KF12:
  \textit{dotted-grey-hatched region}).  The light gray line shows the
  ionizing emissivity model from HM12 constructed to match the WMAP
  $\tau$'s (and the models from KF12 are similar to this [see the
    lower panel of Figure 10 from KF12]).  (\textit{upper right})
  Identical to the upper-left panel except including the optical depth
  constraints from Planck (PC15: \textit{red-shaded region}).
  (\textit{lower left}) Constraints on the evolution of the ionizing
  emissivity (\textit{orange-shaded and green-shaded regions})
  assuming a completion of reionization at $z=5.9$-6.5, the Planck
  Thomson optical depths, and various constraints on $Q_{HII}$ using
  the prevalence of Ly$\alpha$ emission in candidate $z\sim7$-8
  galaxies.  (\textit{lower right}) Similar to the lower left panel,
  but using the $Q_{HII}$ constraints from S14 (\textit{red-shaded
    region}) and also showing the 95\% confidence intervals.  The 68\%
  and 95\% confidence intervals featured in this panel are also
  expressed in tabular form in Table~\ref{tab:backz}.\label{fig:back}}
\end{figure*}

In the present analysis, we consider constraints at four different
points in the reionization history of the universe.  First of all,
there are useful constraints on the end of cosmic reionization at $z =
5.0$-6.5.  The most important constraints make use of the
Gunn-Peterson optical depth measured from bright $z\sim6$ quasars (Fan
et al. 2006a,b; Becker et al.\ 2001) and also by looking at the
distribution of dark gaps in $z = 6$ quasar spectra (McGreer et
al.\ 2015) which suggest that cosmic reionization is complete by
$z=5.9$.  By contrast, the presence of damping absorption wings in the
spectra of three $z=6.2$-6.4 quasars studied by Schroeder et
al.\ (2013) strongly suggest that cosmic reionization is not complete
by $z=6.4$.  These results clearly indicate that cosmic reionization
is completed between $z=6.5$ and $z=5.9$.

Second, we consider the constraints on the reionization history of the
universe that come from the prevalence of Ly$\alpha$ emission in the
UV spectra of $z\sim6$-8 galaxies.  Assuming that the prevalence of
Ly$\alpha$ emission in star-forming galaxies at $z\sim6.5$-8 is a
simple extrapolation of the observed prevalence at $z\sim4$-6 and any
departures from these trends are due to an increasingly neutral IGM at
$z\sim7$-8, one can use this technique to quantify the filling factor
of ionized hydrogen at $z\sim6.5$-8 (Santos 2004; Malhotra \& Rhoads
2004; McQuinn et al. 2007; Mesinger \& Furlanetto 2008; Stark et
al. 2010, 2011; Fontana et al.\ 2010; Dijkstra et al.\ 2011;
Pentericci et al.\ 2011; Ono et al. 2012; Treu et al. 2013; Caruana et
al. 2012, 2014; Tilvi et al.\ 2014; but see also Bolton \& Haehnelt
2013; Mesinger et al.\ 2015).  Here, we will make use of the recent
constraints on $Q_{HII}$ from S14, i.e., $Q_{HII} (z=7) =
0.66_{-0.12}^{+0.09}$ and $Q_{HII} (z=8)<0.35$, using the McQuinn et
al. (2007) models, but also briefly consider the impact of an
alternate set of constraints.

Of course, the real uncertainties on $Q_{HII}$ are likely larger than
the formal uncertainties quoted by S14, as can be seen by comparing
the different estimates for $Q_{HII}$ quoted by S14 based on the same set
of observations or using the results from other studies (Pentericci et
al. 2014; Tilvi et al. 2014).  Therefore, we take the formal $1\sigma$
uncertainty in $Q_{HII}$ to be $\pm$0.15, such that $Q_{HII}
(z=7)=0.66\pm0.15$.  Consistent with the constraints from S14, we
suppose that $Q_{HII}(z=8)$ is equal to $0.15\pm0.20$.  However, we
recognize that the constraints we use are just estimates and the true
values could be different if the assumptions used in deriving these
fractions are not correct (e.g., regarding the velocity offset for
Ly$\alpha$ line or the photoionization rate $\Gamma(z)$: Mesinger et
al. 2015).

Third, we consider the constraints on the cosmic ionizing emissivity
that come from the Thomson optical depth $\tau=0.066\pm0.013$ measured
from the three-year Planck results (PC15).  While this optical depth
measurement does not provide any information on the filling factor of
ionized hydrogen at a specific time, it does provide a powerful
constraint on the integrated path length of ionized hydrogen to the
last-scattering surface.

Finally, we require that the ionizing emissivity extrapolated to
$z=4.75$ be consistent with the $10^{50.99\pm0.45}$ s$^{-1}$
Mpc$^{-3}$ measurement derived by Becker \& Bolton (2013: BB13) based
on observations of the Ly$\alpha$ forest by considering a wide variety
of systematics on this measurement.  Observations of the Ly$\alpha$
forest allow for constraints on the emissivity through the impact of
the photoionization rate $\Gamma(z)$ on the prevalence and statistics
of Ly$\alpha$-forest clouds (and also from constraints on the
mean-free path $\lambda_{mfp}$ using similar observations).

We summarize all four of these key constraints in
Table~\ref{tab:obsconst} and Figure~\ref{fig:obs}, along with other
constraints that have been derived in the literature.  We refer
interested readers to R13 and R15 for a comprehensive summary of these
constraints.

\section{Modeling the Evolution of the Filling Factor of Ionized Hydrogen $Q_{HII}$}

Here we describe the simple evolutionary models we consider for the
evolution of the cosmic ionizing emissivity and which we will compare
against observational constraints on the ionization state of the
universe.

\begin{deluxetable}{ccccc}
\tabletypesize{\footnotesize}
\tablecaption{68\% and 95\% Confidence Intervals on the Inferred Ionizing Emissivity versus Redshift for a fiducial clumping factor $C_{HII}$ of 3.\tablenotemark{a}\label{tab:backz}}
\tablehead{
\colhead{} & \multicolumn{4}{c}{$\log_{10} \dot{N}_{\rm ion}$ [s$^{-1}$ Mpc$^{-3}$]}\\
\colhead{} & \multicolumn{2}{c}{Lower Bound} & \multicolumn{2}{c}{Upper Bound} \\
\colhead{Redshift} & \colhead{95\%} & \colhead{68\%} & \colhead{68\%} & \colhead{95\%}}
\startdata
6 & 50.72 & 50.81 & 51.04 & 51.11 \\
7 & 50.69 & 50.73 & 50.85 & 50.90 \\
8 & 50.52 & 50.60 & 50.71 & 50.74 \\
9 & 50.27 & 50.41 & 50.62 & 50.69 \\
10 & 50.01 & 50.21 & 50.56 & 50.66 \\
11\tablenotemark{b} & 49.75 & 50.00 & 50.49 & 50.64 \\
12\tablenotemark{b} & 49.51 & 49.80 & 50.43 & 50.61 \\
13\tablenotemark{b} & 49.24 & 49.60 & 50.36 & 50.59 \\
14\tablenotemark{b} & 48.99 & 49.39 & 50.29 & 50.57 \\
15\tablenotemark{b} & 48.74 & 49.18 & 50.23 & 50.55 
\enddata
\tablenotetext{a}{The results tabulated here are featured in the lower right panel of Figure~\ref{fig:back} and make use of all 5 key observational constraints considered here (Table~\ref{tab:obsconst}).  These results are derived in the context of the simple two-parameter model described in \S3.}
\tablenotetext{b}{Results here more sensitive to $\tau$'s
  measurements from Planck and functional form adopted in modeling the
  ionizing emissivity evolution.}
\end{deluxetable}

\begin{deluxetable}{ccc}
\tablewidth{0cm}
\tabletypesize{\footnotesize}
\tablecaption{Parameterization for the Cosmic Ionizing Emissivity Satisfying the Key Observational Constraints Considered Here\label{tab:back}}
\tablehead{
\colhead{} & \colhead{$\log_{10} \dot{N}_{\rm ion}(z=8)$} & \colhead{$\dndz$}\\
\colhead{$C_{HII}$} & \colhead{[s$^{-1}$ Mpc$^{-3}$]} &  \colhead{}}
\startdata
2 & $50.62_{-0.07}^{+0.05}$ & $-0.15_{-0.11}^{+0.08}$ \\
3 (fiducial)\tablenotemark{*} & $50.67_{-0.08}^{+0.05}$ & $-0.15_{-0.11}^{+0.08}$ \\
3 (flat prior\tablenotemark{**})\tablenotemark{*} & $50.65_{-0.09}^{+0.06}$ & $-0.19_{-0.11}^{+0.09}$ \\
5 & $50.75_{-0.09}^{+0.06}$ & $-0.14_{-0.10}^{+0.08}$ \\
1+43$z^{-1.71}$$^{\dagger}$ & $50.63_{-0.07}^{+0.05}$ & $-0.16_{-0.10}^{+0.07}$\\\\
\multicolumn{3}{c}{Excluding Ly$\alpha$ Prevalence-Type Constraints\tablenotemark{a}}\\
\multicolumn{3}{c}{(i.e., Excluding Constraints 3-4 from Table~\ref{tab:obsconst})}\\
3 & $50.69_{-0.09}^{+0.05}$ & $-0.09_{-0.13}^{+0.06}$ \\\\
\multicolumn{3}{c}{More Highly Ionized Universe at $z>7$ than S14\tablenotemark{b}}\\
3 & $50.72_{-0.03}^{+0.04}$ & $-0.05_{-0.04}^{+0.03}$ \\\\
\multicolumn{3}{c}{Less Highly Ionized Universe at $z\sim 7$ than S14\tablenotemark{c}}\\
3 & $50.61_{-0.11}^{+0.07}$ & $-0.20_{-0.12}^{+0.10}$\\\\
\multicolumn{3}{c}{WMAP $\tau=0.089\pm0.014$, Reionization at $z=5.9$-6.5}\\
3 & $50.71_{-0.05}^{+0.04}$ & $0.00_{-0.06}^{+0.03}$
\enddata
\tablenotetext{*}{For comparison, we note that only using the Planck $\tau=0.066\pm0.013$ optical depth constraint (PC15), requiring that reionization end at $z=5.9$-6.5, and assuming that $\log_{10} \dot{N}_{\rm ion} (z=4.75) = 50.99\pm0.45$ (BB13), we find $\log_{10} \dot{N}_{\rm ion}(z=8)=50.66_{-0.15}^{+0.07}$ and $\dndz=-0.16_{-0.15}^{+0.10}$ for $C_{HII}=3$.}
\tablenotetext{**}{In our fiducial determinations, the regions of
  parameter space are weighted according to $|\nabla\tau|$ (where the
  derivatives are with respect to $\log_{10}\dot{N}_{\rm ion}$ and
  $\dndz$).  Hence, the prior is flat in units of optical depth
  $\tau$.}
\tablenotetext{$\dagger$}{Redshift Dependence found in the hydrodynamical simulations of Pawlik et al.\ (2009).}
\tablenotetext{a}{As illustrated in Appendix B and Figure~\ref{fig:nnion1MI}}
\tablenotetext{b}{$Q_{HII}(z=7)=0.70\pm0.15$, $Q_{HII}(z=8)=0.55\pm0.20$.  Shown in the lower-left panel of Figure~\ref{fig:back} with the green-shaded region.}
\tablenotetext{c}{$Q_{HII}(z=7)=0.49\pm0.15$, $Q_{HII}(z=8)=0.15\pm0.20$.  Shown in the lower-left panel of Figure~\ref{fig:back} with the orange-shaded region.}
\end{deluxetable}

\begin{figure}
\epsscale{1.17}
\plotone{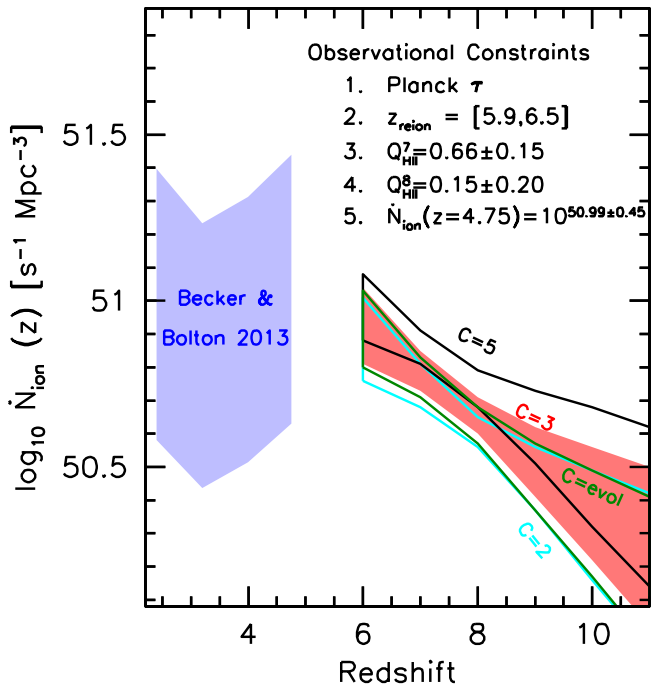}
\caption{Allowed evolution (68\% confidence) of the cosmic ionizing
  emissivity (\textit{region demarcated by the cyan lines},
  \textit{red-shaded region}, and \textit{regions demarcated by the
    black and green lines}) over the redshift range $z=6$ to $z=11$,
  as inferred from the three primary observational constraints
  considered here (Fan et al.\ 2006a; S14; PC15) and assuming the
  clumping factor $C_{HII}$ remains fixed at 2, 3, 5, and evolves as
  $1+43z^{-1.71}$ (as found by the hydrodynamical simulations of
  Pawlik et al.\ 2009), respectively.  The light-blue-shaded region
  indicate the constraints from BB13.  The normalization of the
  derived ionizing emissivity only shows a weak dependence on the
  assumed value of the clumping factor $C_{HII}$, changing by just
  $\sim$30\% for factor-of-2.5 differences in $C_{HII}$ (i.e.,
  $C_{HII}=2$ vs. $C_{HII}=5$).  Strikingly, the evolution inferred
  for the ionizing emissivity, i.e., $\dndz$, is even less sensitive
  to the adopted value for the clumping factor (see also
  Table~\ref{tab:back}).\label{fig:backconc}}
\end{figure}

We model the evolution of the cosmic ionizing emissivity $\dot{N}_{\rm
  ion}(z)$ using just two parameters $A$ and $B$:
\begin{equation}
\dot{N}_{\rm ion}(z)= A\,10^{B(z-8)}
\label{eq:ab}
\end{equation}
where $A=\dot{N}_{\rm ion}(z=8)$ and
$B=\dndz$.  As in previous analyses
(e.g., Madau et al.\ 1999; Bolton \& Haehnelt 2007; KF12), we follow
the evolution of $Q_{HII}$ using the relation
\begin{equation}
\frac{dQ_{HII}}{dt}=\dot{N}_{\rm ion}(z)-\frac{Q_{HII}}{t_{rec}}
\label{eq:m99}
\end{equation}
until $Q_{HII}=1$ when this equation ceases to be valid and the
ionizing emissivity impacts not only atomic-hydrogen gas in the IGM,
but also that in higher-density Ly$\alpha$ clouds.  We take
$Q_{HII}=0$ at $z=25$.  The recombination time $t_{rec}$ is as follows
(KF12):
\begin{equation}
t_{rec}=0.88\textrm{Gyr}\left(\frac{1+z}{7}\right)^{-3}\left(\frac{T_0}{2\times10^4 K}\right)^{-0.7}(C_{HII}/3)^{-1}\label{eq:recomb}
\end{equation}
where $C_{HII}$ is the clumping factor of ionized hydrogen $<n_{HII}
^2>/<n_{HII}>^2$ and $T_0$ is the temperature of the ionizing hydrogen
gas.  We adopt a value of $2\times10^4\,$K for the temperature $T_0$
of the ionizing gas to account for the heating of the gas that occurs
due to the reionization process itself (Hui \& Haiman 2003).

Our calculation of the Thomson optical depths $\tau$ themselves also
follow familiar expressions from previous analyses (e.g., KF12):
\begin{equation}
\tau=\int_0 ^{\infty}dz\frac{c(1+z)^2}{H(z)}Q_{HII}(z)\sigma_T\bar{n}_H(1+\eta\,Y/4X)
\label{eq:tau}
\end{equation}
where $\sigma_T$ is Thomson cross section and $X$ and $Y$ is the
primordial mass fraction of hydrogen and helium.  Following KF12, we
assume that helium is singly-ionized at $z>4$ ($\eta=1$) and
doubly-ionized at $z<4$ ($\eta=2$).

\begin{figure*}
\epsscale{0.7}
\plotone{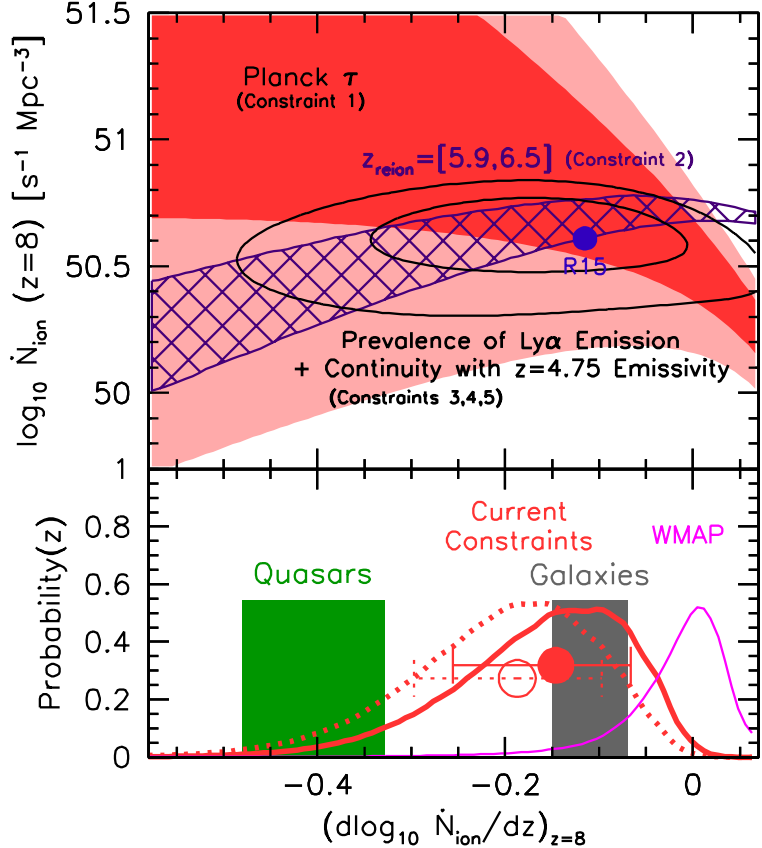}
\caption{(\textit{upper}) Observational constraints on the ionizing
  emissivity $\dot{N}_{\rm ion} (z=8)$ at $z\sim8$ and the evolution
  in this emissivity per unit redshift $\dndz$ (for $C_{HII}=3$).
  68\% and 95\% confidence intervals based on the Planck Thomson
  optical-depth constraints (\textit{red and light-red-shaded
    regions}) and combining the prevalence of Ly$\alpha$ emission in
  $z\sim7$-8 galaxies as found by S14 and continuity with the ionizing
  emissivity at $z=4.75$ as derived by Becker \& Bolton (2013:
  \textit{black lines}).  The purple lines bracket the allowed
  parameters assuming reionization is completed between $z=5.9$ and
  $z=6.5$ (i.e., when $Q_{HII}$ reaches 1 using Eq.~\ref{eq:m99}).
  The blue circle gives the equivalent evolution in the ionizing
  emissivity from the R15 models.  (\textit{lower}) The relative
  likelihood of different rates of evolution in the emissivity,
  $\dndz$, derived by marginalizing over $\dot{N}_{\rm ion}(z=8)$,
  along with the maximum-likelihood value and $1\sigma$ uncertainties
  (\textit{red circle and $1\sigma$ uncertainties}).  The dotted red
  line and open circle show the equivalent likelihoods, for a flat
  prior in $\dot{N}_{\rm ion}(z=8)$ and $\dndz$ (rather than in
  $\tau$: \S4.1), respectively.  Also presented (\textit{magenta
    line}) is the equivalent likelihood using the WMAP optical depths
  and assuming reionization finishes between $z=5.9$ and $z=6.5$ (as
  in the upper left panel of Figure~\ref{fig:back}).  The expected
  redshift dependence of the ionizing emissivity for galaxies and
  quasars is also shown with the shaded grey and green regions,
  respectively.\label{fig:nion_contour}}
\end{figure*}

\section{Results}

\subsection{Quantifying the Redshift Evolution of the Cosmic Ionizing Emissivity}

Here we examine what constraints can be set on the evolution of the
cosmic ionizing emissivity from the key observational constraints we
consider.  We can do this without knowledge of the nature of the
ionizing sources, due to the simplicity of the basic equations that
govern cosmic reionization, i.e., Eq.~(\ref{eq:m99}) and
(\ref{eq:recomb}).  Nothing in these equations requires knowledge of
the nature of the ionizing sources.

To derive constraints on the ionizing emissivity, we consider a full
two-dimensional grid of plausible values of (1) the normalization of
the ionizing emissivity, i.e., $\log_{10} \dot{N}_{\rm ion}(z)_{z=8}$
(from 49.7 to 51.5 s$^{-1}$ Mpc$^{-3}$) and (2) the dependence of this
emissivity on redshift, i.e., $\dndz$ (i.e., from $-$0.6 to 0.1).  For
each choice of the normalization and redshift dependence of the
ionizing emissivity ($A$ and $B$ in Eq.~\ref{eq:ab}), we investigate
whether the assumed emissivity would produce a reionization history
and Thomson optical depth (computed from a reionization history using
Eq.~\ref{eq:tau}) consistent with the key observables we consider
(i.e., constraints 1-5 in Table~\ref{tab:obsconst} and
Figure~\ref{fig:obs}).

We begin by considering the model ionizing emissivities allowed
assuming a redshift-independent clumping factor of 3, but then later
explore what the impact would be of different clumping factors, as
well as considering clumping factors that evolve with redshift.  Our
choice of 3 for the fiducial value of the clumping factor is motivated
by the results of Bolton \& Haehnelt (2007) and Pawlik et al.\ (2009:
see also Finlator et al.\ 2012 and Shull et al.\ 2012).

To illustrate the impact that various observational constraints have
on the evolution of the cosmic ionizing emissivities, we start by
considering only a subset of the available constraints.  More
specifically, we consider the impact of matching both the Thomson
optical depths of different microwave background missions (best
estimate and $\pm1\sigma$) and requiring that reionization be complete
between $z=5.9$ and $z=6.5$ (i.e., when $Q_{HII}$ first reaches 1
using Eq.~\ref{eq:m99}), while taking $C_{HII}=3$.  \S2 describes the
rationale for the $z_{reion}=5.9$-6.5 constraint.  The results are
shown in the top two panels of Figure~\ref{fig:back}.

\begin{figure*}
\epsscale{1.1}
\plotone{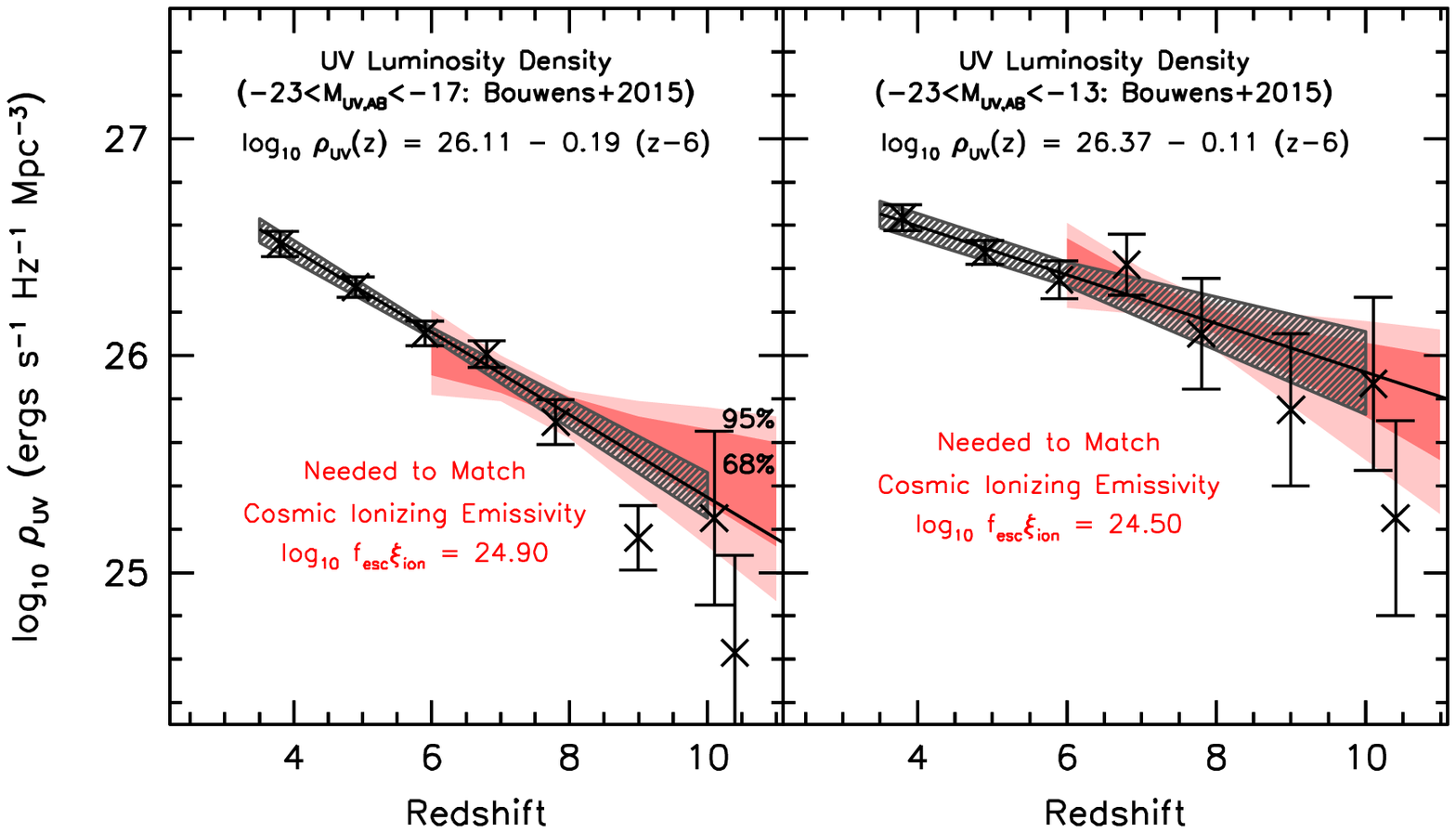}
\caption{(\textit{left}) 68\% and 95\% confidence intervals
  (\textit{red and light red-shaded regions}) on the $UV$ luminosity
  density over the magnitude interval $-23<M_{UV,AB}<-17$ as derived
  by B15 in specific magnitude intervals (\textit{black crosses with
    $1\sigma$ errors}) and based on a fit to the Schechter parameters
  (\textit{black-hatched region}).  These constraints on the $z>6$
  luminosity densities are supplemented by the $z\sim9$ and $z\sim10$
  determinations by Ishigaki et al.\ (2015) and Oesch et al.\ (2015),
  respectively.  Also shown in the panel are the ionizing emissivities
  we infer here ($C_{HII}=3$), offset by a redshift independent
  conversion factor $\log_{10}f_{esc}\xi_{\rm ion}=24.95$.  Even
  though it seems quite clear that $UV$ LF extends faintward of $-17$
  mag (e.g., Alavi et al.\ 2014; Atek et al.\ 2015; B15), we include
  this comparison here to illustrate the similar evolution observed
  when minimal extrapolations are employed.  (\textit{right}) Same as
  left panel, but to a faint-end limit of $-13$ mag and adopting a
  conversion factor of $\log_{10}f_{esc}\xi_{\rm ion}=24.50$.  The
  $UV$ luminosity density integrated down to $-13$ mag likely evolves
  more slowly with redshift than to $-17$ mag, based on the steeper
  shape of the $UV$ LF at high redshift due to an evolution to
  $\alpha$ (e.g., Bouwens et al.\ 2011; McLure et al.\ 2013) and
  possibly $M_{UV}^{*}$ (e.g., Bowler et al.\ 2015; B15).  The $UV$
  luminosity density integrated to $-13$ mag is also more uncertain
  than integrated to $-17$ mag due to the greater extrapolation
  required.  In both the left and right panels, the $UV$ luminosity
  density grows at a similar rate to the inferred ionizing
  emissivities (\S4.2).\label{fig:nnion0}}
\end{figure*}

\begin{figure*}
\epsscale{1.14}
\plotone{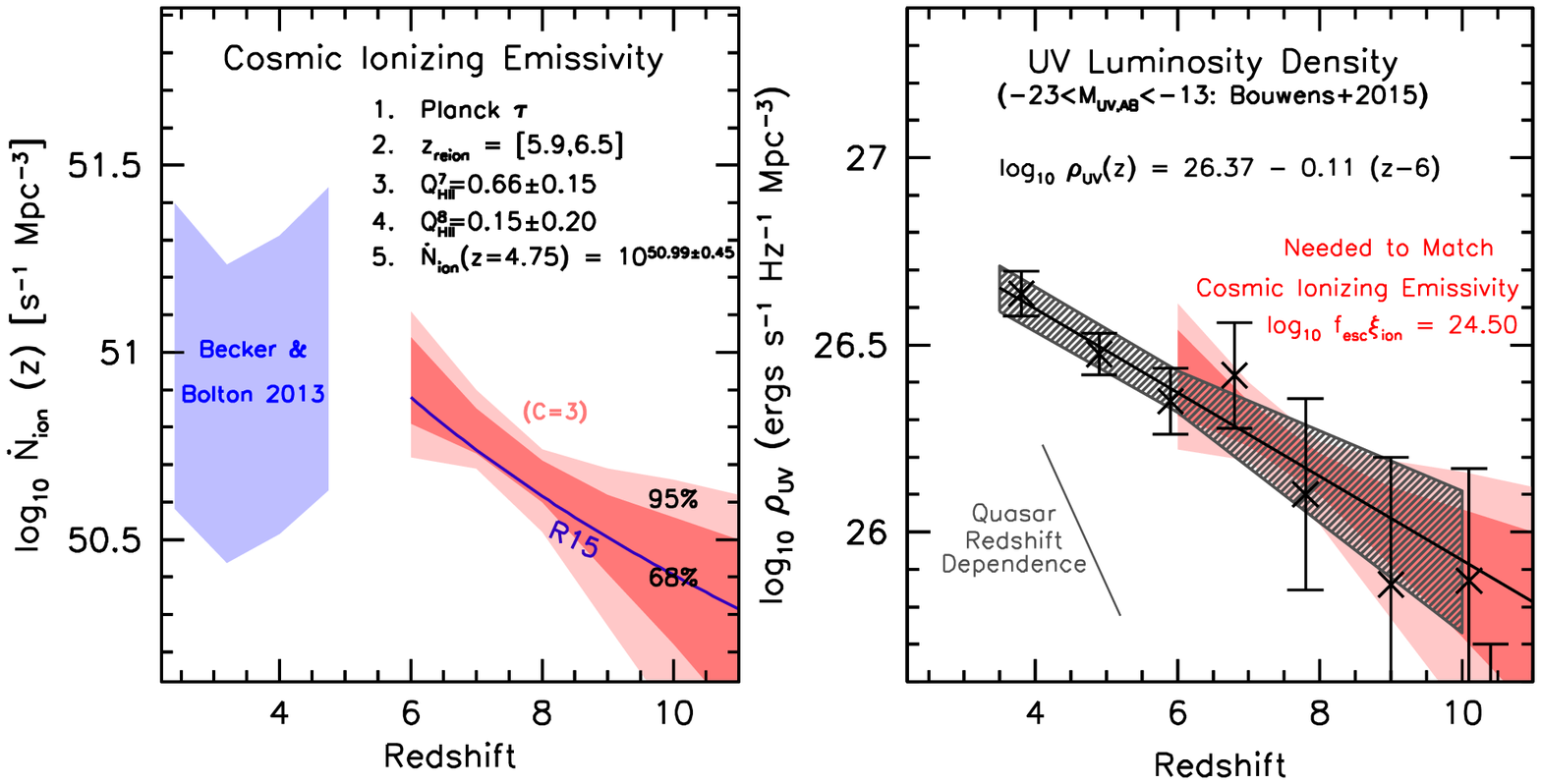}
\caption{(\textit{left}) 68\% and 95\% confidence intervals
  (\textit{red and light red-shaded regions}) on the evolution of the
  cosmic ionizing emissivity (assuming $C_{HII}=3$).  The derived
  evolution presented is the same as in the lower right panel of
  Figure~\ref{fig:back} (see also Table~\ref{tab:backz} for these
  constraints in tabular form).  Constraints from BB13 are indicated
  with the light-blue-shaded region.  The equivalent trend in
  $\dot{N}_{ion}$ derived from R15's model (\textit{blue line}) is
  shown here for context and is included among those ionizing
  emissivity evolution models preferred at 68\% confidence.
  (\textit{right}) 68\% confidence intervals on the $UV-$luminosity
  density over the magnitude interval $-23<M_{UV,AB}<-13$ as derived
  by B15 in specific redshift intervals (\textit{black crosses with
    $1\sigma$ errors}).  See Table~\ref{tab:ldevol} and
  Figure~\ref{fig:nnion0} for the calculated luminosity densities to
  other faint-end limits.  These constraints on the $z>6$ luminosity
  densities are supplemented by the $z\sim9$ and $z\sim10$
  determinations by Ishigaki et al.\ (2015) and Oesch et al.\ (2015),
  respectively, extrapolated to the same limiting luminosity ($-13$
  mag).  The black-hashed region shows the implied evolution of the
  galaxy $UV$-luminosity density, based on our constraints on the
  ionizing emissivity and assuming that galaxies are the source of
  this emissivity with some redshift-independent conversion factor
  $\log_{10}f_{esc}\xi_{\rm ion}=24.50$.  The steep gray line shows
  the redshift dependence one would expect (\textit{with approximately
    the correct normalization [in units of $\dot{N}_{\rm ion}$]
    relative to other quantities presented here}) for the ionizing
  emissivity for quasars using the LEDE-model fit from McGreer et
  al.\ (2013: see \S4.3).  The $UV$ luminosity density and ionizing
  emissivity we infer appear to evolve very similarly
  (\S4.2).\label{fig:nnion1}}
\end{figure*}

As the top two panels illustrate, there are clear differences between
the implied evolution of ionizing sources for WMAP and Planck.
Similar to previous models of the ionizing emissivity which
approximately match the WMAP constraints (e.g., Haardt \& Madau 2012;
KF12), the ionizing emissivity we derive shows no (or a slightly
declining) evolution from $z\sim11$.  For the Planck constraints on
$\tau$, a range of different models are possible, though the ionizing
emissivity in all acceptable models increases with cosmic time.

Interestingly enough, the allowed evolution of the cosmic ionizing
emissivity with redshift shows excellent continuity with the
emissivity inferred at $z=3$-5 from the Ly$\alpha$ forest by Becker \&
Bolton (2013) [BB13].  This suggests that the evolution we have
derived for the ionizing emissivity is plausible.\footnote{While the
  emissivity we derive [extrapolated to $z\sim4$] agrees with BB13, it
  agrees less well with that inferred by KF12 (indicated by the
  \textit{hatched-dotted grey region} in the top left panel of
  Figure~\ref{fig:back}).  This is due to a tension between the
  ionizing emissivity results of KF12 and BB13.  In noting this
  tension, readers should realize that all such determinations of this
  emissivity depend quite sensitively on various model parameters like
  the temperature of the IGM and opacity to ionizing photons.  For
  BB13's determination of the ionizing emissivity, it was possible to
  take advantage of some new measurements of the IGM temperature over
  the redshift range $z=2.0$-4.8 (Becker et al.\ 2011), while also
  including an account for cosmological radiative transfer effects.}

In Figure~\ref{fig:back}, we only present our constraints on the
ionizing emissivity at redshifts $z=6$-12 where changes in the filling
factor of ionized hydrogen will affect the observations we consider.
At redshifts lower than $z\sim6$, the IGM is almost entirely ionized,
and any change in the ionizing emissivity will have little impact on
the $Q_{HII}(z)$'s we consider or the Thomson optical depths.

More detailed constraints on the evolution can be achieved by
examining estimates of the filling factor of ionizing hydrogen
$Q_{HII}$ derived from studies of the prevalence of Ly$\alpha$
emission in $z\sim7$ and $z\sim8$ galaxies and by requiring that the
evolution of the ionizing emissivity be consistent (at $1\sigma$) with
the $z=4.75$ constraints from BB13.  In the lower two panels of
Figure~\ref{fig:back}, the allowed evolution (68\% [left+right] and
95\% [right] confidence) in $\dot{N}_{\rm ion}(z)$ is shown based on
$Q_{HII}$'s derived in S14 (as specified in \S2) and also allowing for
a higher value of $Q_{HII}=0.49\pm0.15$ at $z\sim7$ (e.g., as in
Caruana et al.\ 2014) and $Q_{HII}=0.55\pm0.20$ at $z\sim8$.  The
confidence regions here are derived by calculating the probabilities
based on constraints 1 and 3-5 (Table~\ref{tab:obsconst}) and then
marginalizing over the full two dimensional $\log_{10} \dot{N}_{\rm
  ion}(z=8)$ vs. $\dndz$ space based on the total $\chi^2$ computed
from constraints 1 and 3-5 (Table~\ref{tab:obsconst}).  In
marginalizing over the parameter space, regions where $z_{reion}<5.9$,
$z_{reion}>6.5$, and $\log_{10} \dot{N}_{\rm ion}(z=4.75)>51.44$ are
excluded.  Individual regions of parameter space are weighted
according to the calculated $|\nabla\tau|$ (where the derivatives are
with respect to $\log_{10}\dot{N}_{\rm ion}$ and $\dndz$).  In doing
so, we weight regions of parameter space according to the impact they
have on physical variables like the Thomson optical depth $\tau$
(which approximately varies according to the total output of ionizing
photons at $z\sim8$-15) and give less weight to those regions in
parameter space (i.e., $\dndz<-0.3$) which have less impact on the
ionizing emissivity at $z>>6$.  The 68\% and 95\% confidence intervals
presented in the lower right panel of Figure~\ref{fig:back} are
tabulated in Table~\ref{tab:backz}.

It is clear that if the universe is only 50\% ionized by $z\sim8$ (in
significant contrast to the results of S14), it would imply a much
higher cosmic ionizing emissivity at $z\geq8$.  If we assume that the
universe is measured to be even more ionized at $z\sim7$ and $z\sim8$,
i.e., $Q_{HII}=0.9$, as an even more extreme example (this situation
might arise if use of the Ly$\alpha$ fraction as a probe of $Q_{HII}$
is subject to large systematic errors: e.g., Mesinger et al.\ 2015:
but note also Sobacchi \& Mesinger 2015 and Keating et al.\ 2014), we
derive $\log_{10} \dot{N}_{\rm ion}(z=8)=50.74_{-0.03}^{+0.02}$ and
$\dndz=-0.03_{-0.04}^{+0.03}$.  The inferred cosmic ionizing
emissivity in this case is 0.07 dex higher at $z\sim8$ and 0.31 dex
higher at $z\sim10$ than our fiducial determination.

We found it challenging to reproduce the $<0.3$ filling factors of
ionized hydrogen $Q_{HII}$ at $z\sim7$ found by Ono et al.\ (2012)
within the context of our simple model for the ionizing emissivity, as
it implied optical depths of $<$0.046 (in tension with the Planck
results at 1.7$\sigma$) and also extrapolated to $z=4.75$ implied
ionizing emissivities (i.e., $\log_{10} \dot{N}_{\rm ion} $[s$^{-1}$
  Mpc$^{-3}$]$\sim 51.5$) in excess of that measured by
BB13.\footnote{The range of different constraints on $Q_{HII}$ based
  on the prevalence of Ly$\alpha$ emission in galaxies (often using
  substantially identical observations) illustrate the challenges in
  deriving these $Q_{HII}$ factors, as well as their considerable
  dependence on the simulations used to interpret the available
  observations (and indeed it is clearly non-trivial to adequately
  capture the many relevant physical phenomena that enter into these
  calculations, e.g., growth of structure, star formation, radiative
  transfer of Ly$\alpha$ photons, and patchy reionization in the same
  simulation).}

In Figure~\ref{fig:backconc}, evolution in the ionizing emissivity is
presented, allowing for different values of the clumping factor
$C_{HII}$ and also supposing that the clumping factor evolves in
redshift as $1+43z^{-1.71}$ as found by Pawlik et al.\ (2009) in
sophisticated hydrodynamical simulations.  The primary impact of
changes in the clumping factor is on the overall normalization of
ionizing emissivity $\dot{N}_{\rm ion}$, not its evolution with time.
This can be clearly seen in the maximum likelihood
$\log_{10}\dot{N}_{\rm ion}(z=8)$ and $\dndz$ values presented in
Table~\ref{tab:back} for different values for the clumping parameter.
Interestingly, the difference between the derived evolution of the
ionizing emissivity with $C_{HII}=5$ and $C_{HII}=2$ is just
$\sim$30\%.  This is small (e.g., not a factor of 2.5 as might be
suggested by the ratio of clumping factors) because reionization
appears to be a photon-starved process (Bolton \& Haehnelt 2007).  The
overall insensitivity of estimates of the ionizing emissivity to the
clumping factor $C_{HII}$ is noteworthy.

Our 68\% and 95\% likelihood constraints on the model parameters
$\log_{10} \dot{N}_{\rm ion}(z=8)$ and $\dndz$ are also presented in
Figure~\ref{fig:nion_contour} based on optical depths measured from
Planck, the $Q_{HII}$'s estimated from the prevalence of Ly$\alpha$ in
$z\sim7$-8 galaxies, and requiring that the model emissivity
extrapolated to $z\sim4.75$ matches that derived by BB13.  A flat
prior is assumed in deriving these constraints.  The purple lines
bracket the region allowed for reionization to be completed between
$z=5.9$ and $z=6.5$.

Marginalizing over all values of $\dot{N}_{\rm ion}(z=8)$, we find a
best estimate $\dndz$ of $-0.15_{-0.11}^{+0.08}$ (\textit{red circle}
from the lower panel of Figure~\ref{fig:nion_contour}) based on the
observed constraints.  If we adopt a flat prior in $\dndz$ and
$\dot{N}_{\rm ion}(z=8)$ (instead of our fiducial procedure of taking
the prior to be flat in $\tau$), we find a best estimate $\dndz$ of
$-0.19_{-0.11}^{+0.09}$ (\textit{open red circle} from the lower panel
of Figure~\ref{fig:nion_contour}).

It is worth emphasizing that the evolution we derive for the ionizing
emissivity $\dndz$ is only moderately affected by our use of
constraints that depend on the prevalence of Ly$\alpha$ in $z\sim7$
and $z\sim8$ galaxies (constraints 3-4 in Table~\ref{tab:obsconst}).
Excluding these constraints, we find a best value for $\dndz$ of
$-0.09_{-0.13}^{+0.06}$, $\lesssim$0.5$\sigma$ different from our
fiducial determination.  The left panel of Figure~\ref{fig:nnion1MI}
from Appendix B explicitly illustrates the evolution in the emissivity
we would infer doing the analysis in this manner.

\begin{figure*}
\epsscale{0.85}
\plotone{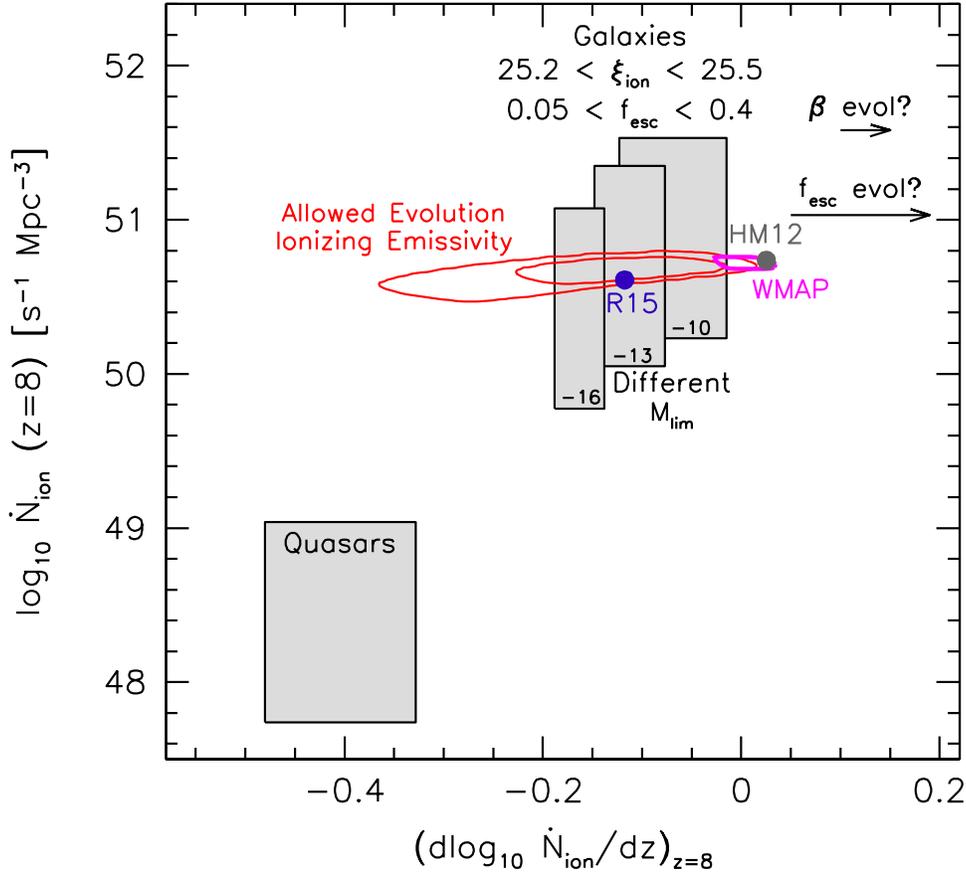}
\caption{68\% and 95\% confidence intervals on parameters describing
  the cosmic ionizing emissivity (assuming $C_{HII}=3$: \textit{red
    contours}) and a comparison to the emissivity expected
  (\textit{gray boxes} in the upper middle) for galaxies to three
  different faint-end cut-offs to the LF ($-10$, $-13$, and $-16$ mag)
  where the Lyman-continuum photon production efficiency $\xi_{\rm
    ion}$ and $f_{esc}$ plausibly have values over the wide range:
  $10^{25.2}$ to $10^{25.5}$ s$^{-1}/($ergs s$^{-1}$Hz$^{-1}$) and
  0.05 to 0.40 based on the observations (e.g., Siana et al.\ 2010,
  2015; Vanzella et al.\ 2012; Mostardi et al.\ 2013; Cooke et
  al.\ 2014; Bouwens et al.\ 2014).  The arrows indicate how potential
  evolution in the $UV$-continuum slopes $\beta$'s or $f_{esc}$ of
  galaxies (as d$\beta/dz\sim-0.04$ [predicted by Wilkins et al.\ 2013
    and consistent with the evolution observed by Bouwens et
    al.\ 2014] or $f_{esc}\propto(1+z)^{3.4}$: required by HM12 to
  match the WMAP $\tau$ measurements) would affect the evolution of
  $\dot{N}_{\rm ion}$.  The gray box to the lower left shows the
  expectations for quasars using the Willott et al.\ (2010) results to
  set the normalization of the ionizing emissivity $\dot{N}_{\rm
    ion}(z)$ (after correcting their results upwards by 0.2 dex to
  account for a possibly steeper faint-end slope $\alpha\sim-2$ versus
  the $\alpha=-1.5$ Willott et al.\ 2010 adopt) and the redshift
  dependence of the emissivity from the McGreer et al.\ (2013) LEDE
  and modified LEDE models.  The 68\% likelihood contours preferred
  based on the WMAP opticle depths and assuming reionization finishes
  at $z=5.9$-6.5 are shown in magenta.  The small size of the WMAP
  contours is an artifact of the sensitivity of the computed optical
  depths $\tau$ to small changes in $\dndz$ when $\dndz$ approaches 0
  (since such implies a constant $\rho_{UV}$ to arbitrarily high
  redshift).  The gray and blue circles give the equivalent parameters
  from the HM12 and R15 models, respectively.  It is clear that
  galaxies should be successful at producing the inferred ionizing
  emissivity for a variety of plausible values for $f_{esc}$,
  $\xi_{\rm ion}$, and $M_{lim}$.  It is also clear from this figure
  that quasars/AGN seem quite unlikely to be the source of this
  emissivity (see also HM12).\label{fig:nnion2}}
\end{figure*}

\subsection{Galaxies as the Primary Source of the Cosmic Ionizing Emissivity?}

We now consider whether galaxies could plausibly be the primary source
of the inferred cosmic ionizing emissivity. To explore this question,
we want to compare both the evolution of the ionizing emissivity and
the evolution of the galaxy UV-luminosity density.

To do this, we first examine the observed UV luminosity density from
the B15 luminosity function (LF) study to observed magnitude limit
$-17$ mag from current observations (see Figures 15, 18 and 19 from
B15).  The left panel of Figure~\ref{fig:nnion0} shows the observed
results for the UV luminosity density $\rho_{UV}$ from B15 and from
other sources (e.g., Oesch et al.\ 2015; Ishigaki et al.\ 2015).

It is clear given the steep slope at $-$17 mag and the lack of any
indication of a cut-off from Figures 15 and 19 in B15 for example (see
also Schenker et al. 2013; McLure et al. 2013; Alavi et al. 2014;
Barone-Nugent et al. 2015), as well as early results from magnified
sources found in lensing clusters (e.g., Atek et al 2015) that fainter
galaxies must contribute substantially to the total UV radiation from
galaxies (e.g., Yan \& Windhorst 2004; Beckwith et al.\ 2006; Bouwens
et al.\ 2007; Reddy \& Steidel 2009).  While we can make plausible
extrapolations based on the derived Schechter parameters, the question
arises as to the appropriate limit.

As others have done (e.g., R13, R15) we adopt a fiducial luminosity of
$-$13 mag down to which galaxies are typically assumed to be able to
form efficiently.  Faintward of $-$13 mag, galaxy formation may be
suppressed due to the inefficient gas cooling onto low-mass halos
(e.g., Rees \& Ostriker 1977) or due to the difficulties in low-mass
galaxies’ retaining their gas (e.g., Mac Low \& Ferrara 1999; Dijkstra
et al. 2004). The choice of the limiting luminosity to adopt for the
LF, e.g., $M_{lim}$, is also motivated from sophisticated
hydrodynamical simulations (e.g., O'Shea et al. 2015) or from fits to
the faintest points in the LF (e.g., Mu{\~n}oz \& Loeb 2011;
Barone-Nugent et al.  2015).  Some guidance can also be obtained by
attempting to match the redshift dependence inferred for ionizing
emissivity $\dot{N}_{\rm ion}(z)$ with $\rho_{UV}$ (e.g., Bouwens et
al. 2012a; R13), i.e., compare the inferred $\dndz$'s in
Table~\ref{tab:back} with the $d\rho_{UV}/dz$'s in
Table~\ref{tab:ldevol}, but this presupposes that galaxies are the
only source of the ionizing UV radiation and so it is considered indicative.

With our adopted limiting magnitude of $-13$ mag, we estimate the
UV-luminosity density $\rho_{UV}$ implied by the B15 LF results, by
marginalizing over the full likelihood distribution in $M^*$,
$\phi^*$, and $\alpha$ and computing both the mean and $1\sigma$ error
from the resultant likelihood distribution for $\rho_{UV}$ (similar to
Figure 3 from R13).

The estimated UV-luminosity densities $\rho_{UV}$ to $-$13 mag are
shown in the right panel of Figure~\ref{fig:nnion0}.  The shallower
redshift dependence of the result is due to the inclusion of much
lower-luminosity galaxies in the calculated luminosity densities and
the strong evidence that the $UV$ luminosity density evolves faster at
the bright end than the faint end (e.g., Yoshida et al.\ 2006; Bouwens
et al.\ 2006, 2008; B15).  The most important factor in this
differential evolution is the flattening of the faint-end slope
$\alpha$ to the $UV$ LF with cosmic time (e.g., Bouwens et al.\ 2011;
Schenker et al.\ 2013; McLure et al.\ 2013).  For context, the
luminosity densities $\rho_{UV}$ implied by the B15 Schechter
parameters to many different limiting luminosities and as a function
of redshift are presented in Table~\ref{tab:ldevol}.  The red regions
on the two panels of Figure~\ref{fig:nnion0} will be discussed below
following the discussion of Figure~\ref{fig:nnion1}.

We are now positioned to explore both the evolution of the ionizing
emissivity and the evolution of the galaxy UV-luminosity density, as
shown in the left and right panels of Figure~\ref{fig:nnion1},
respectively, with the vertical axes chosen so that the inferred
ionizing emissivity and luminosity density maximally overlap, i.e.,
$\dot{N}_{\rm ion}(z)=f_{esc}\xi_{\rm ion}\rho_{UV}$.

As is apparent from Figure~\ref{fig:nnion1}, the best-fit evolution in
the ionizing emissivity $\dndz$ of $-0.15_{-0.11}^{+0.08}$ (or
$-0.19_{-0.11}^{+0.09}$ adopting a flat prior in $\dndz$ and
$\log_{10} \dot{N}_{\rm ion}(z=8)$) is in excellent agreement with the
best-fit evolution in the $UV$ luminosity density to $-13$ mag, i.e.,
$d\log_{10} \rho_{UV}/dz =-0.11\pm0.04$.  While the uncertainties are
still large, this is suggestive that early star-forming galaxies
provide the ionizing photons needed to reionize the universe.

We emphasize that the present conclusions do not significantly depend
on our use of the observational constraints based on the
prevalence of Ly$\alpha$ emission from $z\sim7$ and $z\sim8$ galaxies.
As we discuss in \S4.1, if we exclude these constraints, the $\dndz$
value we derive ($-0.09_{-0.13}^{+0.06}$) is almost identical to the
best-fit evolution we find in the $UV$ luminosity density.  See
Appendix B and Fig.~\ref{fig:nnion1MI} where this point is illustrated
explicitly.

It is interesting also to consider the offset where the ionizing
emissivity and $UV$ luminosity density maximally overlap.  The best
overlap occurs adopting an offset of $10^{24.50}$ s$^{-1}/($ergs
s$^{-1}$Hz$^{-1})\,(=f_{esc}\xi_{\rm ion})$ for a faint-end limiting
luminosity of $-13$ mag (similar to what has been proposed in many
other studies: e.g., KF12, R13).  This conversion factor has an
uncertainty of at least 0.10 dex, given the uncertainties on both the
normalization of the ionizing emissivity at $z\sim8$ (typically
$\sim$0.08 dex: Table~\ref{tab:back}) and the $\gtrsim$0.05 dex
uncertainties on the $UV$ luminosity density integrated to
$\lesssim-13$ mag (see Table~\ref{tab:ldevol}).  This is one of the
first analyses to estimate an uncertainty on this conversion factor
using observations that directly concern the $z>6$ universe (see also
Mitra et al.\ 2013).  The evolution of the ionizing emissivity is also
shown in relation to the $\rho_{UV}$ for the two magnitude limits in
Figure~\ref{fig:nnion0}, using an appropriate offset.  It is quite
clear that the two quantities seem to evolve in a similar manner even
to the bright limit.

Is this multiplicative offset plausible if star-forming galaxies are
to be the source of the cosmic ionizing emissivity?  Given that it can
be expressed as the product of two factors $\xi_{\rm ion}$, the
production efficiency of Lyman-continuum photons per unit $UV$
luminosity, and the escape fraction $f_{esc}$ with plausible values in
the range $10^{25.2}$ to $10^{25.5}$ s$^{-1}/($ergs s$^{-1}$Hz$^{-1}$)
and 0.05 to 0.40 based on observations of $z\sim2$-4 galaxies
(Vanzella et al.\ 2012; Mostardi et al.\ 2013; R13; Duncan \&
Conselice 2015), the observed offset between the inferred ionizing
emissivity and UV luminosity density is certainly within the range
allowed by these value, i.e., $10^{23.9}$ to $10^{25.1}$
s$^{-1}/($ergs s$^{-1}$Hz$^{-1}$).  This is the case for all three
limiting luminosities $M_{lim}$ presented in Figure~\ref{fig:nnion2}
($-10\,\,$mag, $-13\,\,$mag, and $-16\,\,$mag).

In distinct contrast to the situation faced in interpreting the WMAP
$\tau$'s, no evolution in the escape fraction is required to match the
ionizing emissivity model we infer, as had been earlier considered by
HM12 or KF12.  Neither is evolution in the $UV$-continuum slopes
$\beta$ required, despite the apparent trend towards bluer $\beta$'s
at high redshift (e.g., d$\beta/dz\sim-0.10\pm0.05$: Bouwens et
al.\ 2014: see also Bouwens et al.\ 2012; Kurczynski et al.\ 2014;
Finkelstein et al.\ 2012a; Wilkins et al.\ 2011) or as expected from
simple theoretical models (e.g., d$\beta/dz\sim-0.04$ [Wilkins et
  al.\ 2013] or d$\beta/dz\sim-0.1$ [Finlator et al.\ 2011]).  A mild
evolution in $\beta$ is not inconsistent with our constraints on the
ionizing emissivity.

\begin{deluxetable}{cc}
\tablewidth{0cm}
\tabletypesize{\footnotesize}
\tablecaption{$UV$ Luminosity Density derived from the B15 LF parameters integrated to different faint-end cut-offs $M_{lim}$ to the $UV$ LF.\label{tab:ldevol}}
\tablehead{
\colhead{$M_{lim}$} & \colhead{$\log_{10} \rho_{UV}$ [ergs s$^{-1}$Hz$^{-1}$ Mpc$^{-3}$]}}\\
\startdata
$-10$  & $(26.47 \pm 0.08) - (0.07\pm0.05) (z-6)$ \\
$-13$ (fiducial) & $(26.37\pm0.05) - (0.11\pm0.04) (z-6)$\\
$-16$  & $(26.20 \pm 0.03) - (0.16\pm0.03) (z-6)$ \\
$-17$\tablenotemark{a}  & $(26.11 \pm 0.03) - (0.19\pm0.02) (z-6)$ \\
$-17.7$\tablenotemark{a} & $(26.02 \pm 0.03) - (0.21\pm0.02) (z-6)$
\enddata

\tablenotetext{a}{The evolution of the $UV$ luminosity density to
  $-17$ and $-17.7$ mag (as used by Bouwens et al.\ 2011; Ellis et
  al.\ 2013; Oesch et al.\ 2013, 2014, 2015; McLeod et al.\ 2015) is
  included for contrast with the evolution seen to much fainter limits
  and also for completeness.  The steep slope of the $UV$ LF at the
  $-17$ mag limit of deep searches for faint galaxies at $z\sim7$-8
  suggests that the cut-off is fainter than $-17$ mag (e.g., B15;
  Schenker et al.\ 2013; McLure et al.\ 2013).  Recent results from
  $z\sim7$ and $z\sim8$ LFs derived from magnified sources behind
  lensing clusters also indicate a steep slope, and extends the limit
  at $z\sim 7$ to $-$15.5 mag (e.g., Atek et al.\ 2015).}
\end{deluxetable}

Given the likely connection between the ionizing emissivities we infer
and galaxy $UV$ luminosity density at $z>6$, what are physically
plausible values for $\xi_{\rm ion}$ and $f_{esc}$ that we would
infer?  The relevant value of the $UV$-continuum slopes $\beta$ of
galaxies to estimate their ionizing emissivity contribution is
$\sim-2.3$, as most of the luminosity density at $z\geq6$ occurs in
lower luminosity galaxies and the median $\beta$ for faint $z\sim7$-8
galaxies is $\sim-2.3$ (e.g., Bouwens et al.\ 2014).  Using a similar
approach to R13, one can show that $\xi_{\rm ion}$ in such galaxies
has an approximate value of $10^{25.46}$ s$^{-1}/($ergs
s$^{-1}$Hz$^{-1}$) for $\xi_{\rm ion}$ (Appendix A), similar to the
$\xi_{\rm ion}$ advocated by Topping \& Shull (2015).  For this value
of $\xi_{\rm ion}$, the escape fraction $f_{esc}$ is
0.11$_{-0.02}^{+0.03}$ for the observed conversion factor $10^{24.50}$
s$^{-1}/($ergs s$^{-1}$Hz$^{-1}$).  Alternatively, if we take
$\xi_{\rm ion}$ to be equal $10^{25.2}$ s$^{-1}/($ergs
s$^{-1}$Hz$^{-1}$) as R13 adopted to match with the $\beta$
measurements of Dunlop et al.\ (2013), the relevant value of $f_{esc}$
is 0.20$_{-0.04}^{+0.05}$ (essentially identical to R13's adopted
value of 0.2).

Clearly, many degenerate combinations of $f_{esc}$, $\xi_{\rm ion}$,
and $M_{lim}$ can be successful in producing the same ionizing
emissivity.  For variable $M_{lim}$, the present constraint on
$f_{esc}\xi_{\rm ion}$ can be generalized to the following constraint
on these three parameters:
\begin{equation}
f_{esc}\xi_{\rm ion} f_{corr}(M_{lim}) = 10^{24.50} \textrm{s}^{-1}/(\textrm{ergs s}^{-1}\textrm{Hz}^{-1})
\label{eq:convf}
\end{equation}
where the added term $f_{corr} (M_{lim}) =
10^{0.02+0.078(M_{lim}+13)-0.0088(M_{lim}+13)^2}$ corrects
$\rho_{UV}(z=8)$ derived to a faint-end limit of $M_{lim}=-13$ mag to
account for different $M_{lim}$'s.\footnote{In deriving this
  correction factor, we made use of the following relationship between
  $\rho_{UV}(z=8)$ and the faint-end cut-off $M_{lim}$ to the LF:
  $\log_{10} \rho_{UV}(z=8) = (26.17\pm0.09) + (0.08\pm0.02)
  (M_{lim}+13) - (0.009\pm0.008) (M_{lim}+13)^2$.  This relationship
  can be derived by fitting to the results in Table~\ref{tab:ldevol}
  and is accurate to 5\%.}  For clumping factors $C_{HII}$ not equal
to our fiducial value of 3, the left-hand side of the above equation
should be multipled by $(C_{HII}/3)^{-0.3}$ based on the results
presented in Table~\ref{tab:back} (the $(C_{HII}/3)^{-0.3}$ scaling is
approximately valid for $C_{HII}<10$).  The factor $10^{24.50}$
s$^{-1}/($ergs s$^{-1}$Hz$^{-1}$) has an uncertainty of $\sim$0.1 dex.

Eq.~(\ref{eq:convf}) can be manipulated to allow for an estimate of
$f_{esc}$ given assumed values for the other parameters $M_{lim}$,
$\xi_{ion}$, and $C_{HII}$, assuming that galaxies reionize the
universe: 
\begin{equation}
f_{esc} \approx \xi_{\rm ion}^{-1} f_{corr}^{-1}(M_{lim}) (C_{HII}/3)^{0.3} 10^{24.50} \textrm{s}^{-1}/(\textrm{ergs s}^{-1}\textrm{Hz}^{-1})
\label{eq:fesc}
\end{equation}
Estimates of $f_{esc}$ for various fiducial choices of $\xi_{ion}$,
$M_{lim}$, and clumping factors $C_{HII}$ are provided in
Table~\ref{tab:fesc}.  This is one of the first analyses to estimate
the uncertainty on the derived value of $f_{esc}$ based on
observational constraints on the ionization state of the $z>6$
universe (see also Mitra et al.\ 2013).

The issue of degeneracies among the parameters $f_{esc}$, $\xi_{\rm
  ion}$, and $M_{lim}$ is discussed extensively in KF12 (see also
Bouwens et al.\ 2012a; Finkelstein et al.\ 2012b; R13) and may not be
easy to resolve based on observations in the immediate future.

\subsection{Quasars as the Primary Source of the Cosmic Ionizing Emissivity?}

We briefly consider whether quasars could be the primary source of the
cosmic ionizing emissivity.  Despite their relative scarcity in the
$z>4$ universe, quasars potentially can contribute quite substantially
to the inferred emissivity due to the hardness of their spectrum and a
much higher escape fraction ($f_{esc}=1$?: e.g., Loeb \& Barkana
2001).

While a fraction of the emissivity from quasars at $z>4$ is expected
to originate from the most luminous sources, a potentially large
fraction of their contribution could originate at much lower
luminosities, and therefore it is important to have reasonable
constraints on both the volume densities of faint quasars (including
AGN) and their faint-end slopes to effectively estimate the ionizing
emissivity they produce.

Perhaps the deepest, wide-area probes of the $z>4$ quasar LFs are
provided by Willott et al.\ (2009, 2010) and McGreer et al.\ (2013).
The deepest part of Willott et al.\ (2009, 2010) searches for faint
($>-22$ mag) $z\sim6$ QSOs over 4.5 deg$^2$, while McGreer et
al.\ (2013) leveraged the deep observations over a 235 deg$^2$ region
in SDSS Stripe 82 to probe the prevalence of moderately faint ($>-24$
mag) $z\sim5$ qusars.  Willott et al.\ (2009) identified 1 very faint
$z=6.01$ quasar over the CFHTLS Deep/SXDS, while McGreer et
al.\ (2013) identified some $\sim$70 faint quasars in their search, 29
with absolute magnitudes faintward of $-25$ mag.

\begin{deluxetable*}{ccccccc}
\tablewidth{0cm}
\tabletypesize{\footnotesize}
\tablecaption{Required Values of $f_{esc}$ for different $M_{lim}$, $\xi_{ion}$, and clumping factors $C_{HII}$ assuming that galaxies drive the reionization of the universe.\tablenotemark{a}\label{tab:fesc}}
\tablehead{
\colhead{} & \multicolumn{6}{c}{Required $f_{esc}$} \\
\colhead{} & \multicolumn{3}{c}{$\xi_{ion}=10^{25.46}$ s$^{-1}/($ergs
s$^{-1}$Hz$^{-1}$)} & \multicolumn{3}{c}{$\xi_{ion}=10^{25.2}$ s$^{-1}/($ergs
s$^{-1}$Hz$^{-1}$)\tablenotemark{b}} \\
\colhead{$C_{HII}$} & \colhead{$M_{lim}=-17$} & \colhead{$M_{lim}=-13$} & \colhead{$M_{lim}=-10$} & \colhead{$M_{lim}=-17$} & \colhead{$M_{lim}=-13$} & \colhead{$M_{lim}=-10$}}
\startdata
2 & 0.26$_{-0.05}^{+0.07}$ & 0.10$_{-0.02}^{+0.03}$ & 0.06$_{-0.01}^{+0.02}$ & 0.46$_{-0.10}^{+0.12}$ & 0.18$_{-0.04}^{+0.05}$ & 0.12$_{-0.02}^{+0.03}$\\
3 & 0.29$_{-0.06}^{+0.07}$ & 0.11$_{-0.02}^{+0.03}$ & 0.07$_{-0.01}^{+0.02}$ & 0.52$_{-0.11}^{+0.14}$ & 0.20$_{-0.04}^{+0.05}$\tablenotemark{c} & 0.13$_{-0.03}^{+0.03}$\\
5 & 0.34$_{-0.07}^{+0.09}$ & 0.13$_{-0.03}^{+0.03}$ & 0.08$_{-0.02}^{+0.02}$ & 0.61$_{-0.13}^{+0.16}$ & 0.23$_{-0.05}^{+0.06}$ & 0.15$_{-0.03}^{+0.04}$\\
1+43$z^{-1.71}$$^{\dagger}$ & 0.27$_{-0.06}^{+0.07}$ & 0.10$_{-0.02}^{+0.03}$ & 0.07$_{-0.01}^{+0.02}$ & 0.49$_{-0.10}^{+0.13}$ & 0.19$_{-0.04}^{+0.05}$ & 0.12$_{-0.03}^{+0.03}$
\enddata
\tablenotetext{$\dagger$}{Redshift Dependence found in the hydrodynamical simulations of Pawlik et al.\ (2009).}
\tablenotetext{a}{These $f_{esc}$ factors can be derived from Eq.~(\ref{eq:fesc}) in \S4.2 of this paper.  Importantly, we can also quote uncertainties on the estimated $f_{esc}$'s, which follow from our $1\sigma$ error estimate ($\sim$0.1 dex) on the conversion factor $10^{24.50} \textrm{s}^{-1}/(\textrm{ergs s}^{-1}\textrm{Hz}^{-1}$) from $UV$ luminosity density $\rho_{UV}$ to the equivalent ionizing emissivity $\dot{N}_{\rm ion}$.  Constraints on $f_{esc}$ are also attempted by KF12 at $z\sim4$ based on the derived $\dot{N}_{\rm ion}$ there (see also Finkelstein et al.\ 2012b for an estimated $f_{esc}$ at $z\sim6$ based on $\dot{N}_{\rm ion}$ from Bolton \& Haehnelt 2007).}
\tablenotetext{b}{Adopted by R13.}
\tablenotetext{c}{In fact, this is the same $f_{esc}$ that R13 and R15 suggest using for the fiducial parameter choices they adopt: $\xi_{ion}=10^{25.2}$ s$^{-1}/($ergs s$^{-1}$Hz$^{-1}$), $M_{lim}=-13$, and $C_{HII}=3$.  However, no attempt was made by R15 to provide a constraint on the uncertainties in this estimate of $f_{esc}$ as derived here.}
\end{deluxetable*}

Given the relatively small numbers of faint quasars identified by
these programs (and other recent searches e.g., Weigel et al.\ 2015,
who find no convincing $z\gtrsim5$ AGN over the Chandra Deep Field
South), the total emissivity of ionizing photons from the quasar
population at $z\sim4$-6 is still quite uncertain.  Nevertheless, a
sufficient number of faint quasars have been found that estimates of
the total emissivity of the population can still be made (but see the
discussion in Giallongo et al.\ 2015).  At this time, the most
reliable estimates can be made using the faint-end slopes $\alpha$ and
evolutionary trends derived by McGreer et al.\ (2013).  McGreer et
al.\ (2013) find a $\log_{10}\phi(z)= \log_{10} \phi_0 - 0.60(z-2.2)$
and $M_*(z) = M_{*,0} - 0.68(z-2.2)$ trend with their LEDE model and a
$\log_{10}\phi(z)= \log_{10} \phi_0 - 0.70(z-2.2)$ and $M_*(z) =
M_{*,0} - 0.55(z-2.2)$ trend with their modified LEDE model.  These
models imply a comoving emissivity $\log_{10} \epsilon(z)$ that
evolves as $-0.33(z-2.2)$ and $-0.48(z-2.2)$.

If we take the ionizing emissivity estimates that Willott et
al.\ (2010) derive from their deep $z\sim6$ search, extrapolate to
$z\sim8$, and adjust their result upwards by 0.2 dex to account for
the steeper faint-end slope of $\alpha=-2$ found at $z\sim5$ by
McGreer et al.\ (2013: versus an assumed faint-end slope of
$\alpha=-1.5$ by Willott et al.\ 2010), the $\log_{10} \dot{N}_{\rm
  ion}(z=8)$ we would derive is 48.5, with a $\dndz$ of $-0.41\pm0.08$
taking the arithmetic mean of the two LEDE models considered by
McGreer et al.\ (2013).  Fan et al.\ (2001) derive $-0.47\pm0.15$
based on the evolution of the space density of bright quasars from
$z=6$ to $z=3$.  If we compare these parameters with the values we
derive for the ionizing emissivity, it is clear that faint quasars do
not appear to come close to providing enough photons to drive the
reionization of the universe.  In addition, the ionizing emissivity
produced by this population shows a redshift dependence that is
slightly steeper than what we require to reproduce the observational
constraints (see e.g. the green-shaded region and the red line at the
bottom of Figure~\ref{fig:nion_contour}).

\begin{figure}
\epsscale{1.19}
\plotone{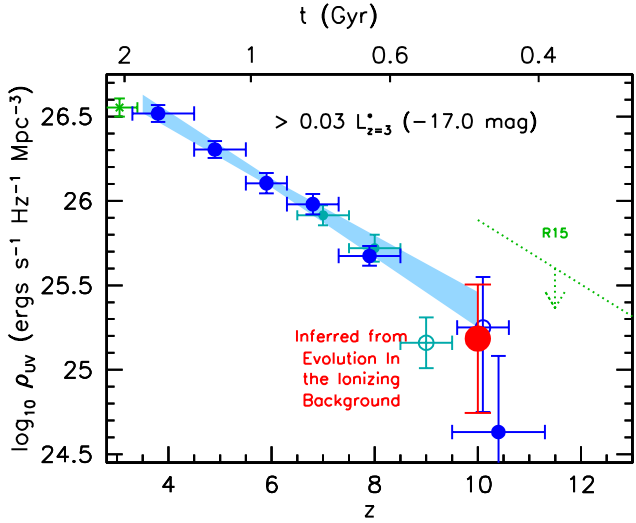}
\caption{Constraints on the $UV$ luminosity density at $z\sim10$ to
  the typical observed limit of $-17$ mag (\textit{red circle}),
  derived from the $z=6$-10 evolution of the ionizing emissivity
  (\S4.4).  The luminosity density results from B15 at $z=6$ are used
  as a baseline.  The derived UV luminosity density at $z=10$ is also
  corrected downward by $-0.32$ dex to account for the different
  evolution in $\rho_{UV}$ expected to $-17$ mag vs. to $-13$ mag (see
  Table~\ref{tab:ldevol}).  For context, estimates of the $UV$
  luminosity density to the same faint-end limit based on the Reddy \&
  Steidel (2009) $z=3$ results (\textit{green cross}), the B15
  $z=4$-10 results (\textit{blue circles}), the McLure et al.\ (2013)
  $z=7$-8 results (\textit{cyan circles}), the Ishigaki et al.\ (2015)
  $z=9$ (\textit{cyan open circles}), and Oesch et al.\ (2015) $z=10$
  results (\textit{blue open circles}) are also shown.  The
  shaded-light-blue curve indicates the evolution in $UV$ luminosity
  density one would expect based on extrapolations of the B15 LF
  parameters to $z>8$.  The dotted green $1\sigma$ upper limit at
  $z=11.5$ shows constraints on the $UV$ luminosity density at $z>10$
  from R15 derived using the recent Planck $\tau$ measurements
  corrected to $-17$ mag based on the empirical scalings presented in
  Table~\ref{tab:ldevol}.
\label{fig:sfz}}
\end{figure}

Is it possible that even fainter quasars/AGN could boost these
luminosity densities?  Given that the faint-end slope $\alpha$ McGreer
et al.\ (2013) find at $z\sim5$ is $-2.03_{-0.14}^{+0.15}$, the
luminosity density will be only logarithmically divergent, which for
an integration to $\sim-22$ mag and $\sim-16$ mag (if such fainter
quasars/AGN exist in large numbers) would only increase the overall
luminosity density by a factor of 1.5 (0.2 dex) and a factor of 4 (0.6
dex).  This suggests that a full consideration of the contribution
from faint quasars can potentially boost the total ionizing emissivity
produced by quasars.  However, even with such steps, quasars appear
quite unlikely to contribute meaningfully to the reionization of the
universe at $z>6$ (but see Giallongo et al.\ 2015).

\subsection{Implications for the UV Luminosity Density at $z\sim10$?}

The current discussion and others (e.g., R15; Mitra et al.\ 2015)
suggest that star-forming galaxies produce the bulk of the ionizing
emissivity.  Given this, we can use current constraints on the
evolution of the ionizing emissivity to provide an estimate of the
$UV$ luminosity density at $z\sim10$.  This is useful since there has
been much discussion about whether the UV luminosity and SFR density
are trending differently with time from $z\sim 10$ to $z\sim 8$ than
at later times (e.g., Oesch et al.\ 2012, 2013, 2014, 2015; Ellis et
al.\ 2013; Coe et al.\ 2015; Ishigaki et al.\ 2015; McLeod et
al.\ 2015).

In the simplest case of no evolution in $f_{esc}$ or $\xi_{\rm ion}$,
evolution in the $UV$ luminosity density would mirror that seen in
ionizing emissivity.  In this case, one would expect a factor of
$10^{4(-0.15_{-0.11}^{+0.08})} = 4_{-2}^{+7}$ increase in the
$UV$ luminosity density from $z\sim10$ to $z\sim6$ integrated to the
faint-end cut-off to the LF $M_{lim}$, which we will take to be $-13$
mag.  However, since the UV luminosity density is measured directly
from observations to about $-17$ AB mag in the reionization epoch, it
is perhaps preferable to evaluate the changes to that well-determined
limit and to correct the derived results from the ionizing emissivity
to $-17$ mag instead of the extrapolated $-13$ mag limit.  Considering
the evolution integrated to $-17$ mag, we would expect d$\log_{10}
\rho_{UV}/dz = -0.19 - (-0.11) = 0.08$ more evolution at the bright
end than the faint-end (using the scalings from
Table~\ref{tab:ldevol}: see also Yoshida et al.\ 2006; Bouwens et
al.\ 2007, 2008 where such differential trends were first discussed).
This would suggest an increase of $8_{-4}^{+15}\times$ in the
luminosity density from $z\sim10$ to $z\sim6$ to $-17$ mag.
Integrating the $z\sim6$ B15 LF to $-17\,\,$mag and accounting for
this evolution, we estimate a $UV$ luminosity density of
$10^{25.19_{-0.44}^{+0.36}}$ ergs s$^{-1}$ Hz$^{-1}$Mpc$^{-3}$ at
$z\sim10$ to $-17\,$mag (Figure~\ref{fig:sfz}).  This estimate should
be regarded as an upper limit, as any expected mild evolution in
$f_{esc}$ or $\xi_{\rm ion}$ (Siana et al.\ 2010; Hayes et al.\ 2011;
Bouwens et al.\ 2014: but see also R13) would likely act to lower the
derived luminosity density at $z\sim10$.

\begin{figure*}
\epsscale{1.15} \plotone{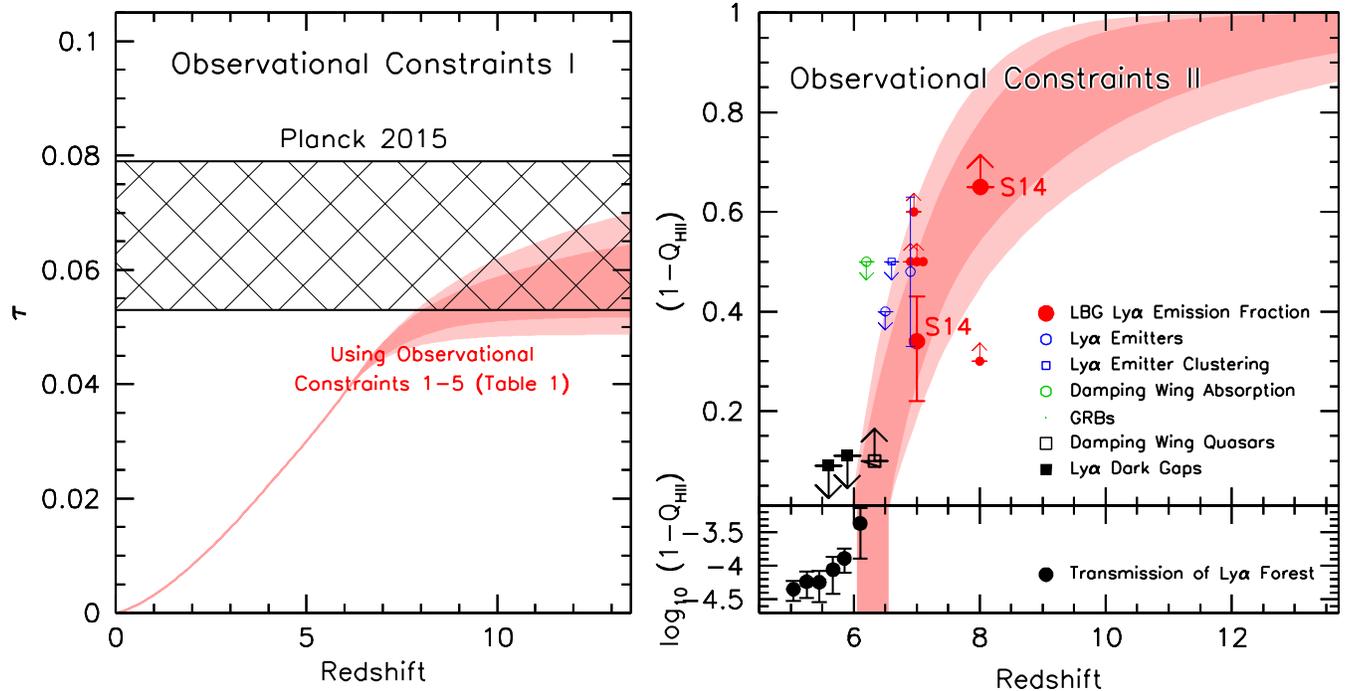}
\caption{Comparison of the key observational constraints considered
  here with the results from the simple two-parameters models for the
  cosmic ionizing emissivity preferred at 68\% and 95\% confidence
  (\S4.5).  (\textit{left}) Shown are the constraints on the Thomson
  optical depth $\tau$ provided by the Planck 3-year results (PC15:
  \textit{cross-hatched black region}).  The red and light-red-shaded
  regions show the range of cumulative Thomson optical depths for our
  models of the ionizing emissivity preferred at 68\% and 95\%
  confidence, respectively (Figure~\ref{fig:nion_contour}), and where
  reionization is complete between $z=5.9$-6.5.  (\textit{right})
  Shown are constraints on the filling factor of ionized hydrogen
  $Q_{HII}$ as a function of redshift.  The constraints are largely as
  compiled by R15 (see Table~\ref{tab:obsconst} of the present
  manuscript) and are based on the Gunn-Peterson optical depths and
  dark-gap statistics measured in $z\sim6$ quasars (Fan et al.\ 2006a;
  McGreer et al.\ 2015: \textit{solid black circles and squares}),
  damping wings measured in $z\sim6.2$-6.4 quasars (Schroeder et
  al.\ 2013: \textit{open black square}) and a $z=6.3$ GRB (Totani et
  al.\ 2006; McQuinn et al.\ 2008: \textit{open green circle}),
  Ly$\alpha$-emitter LFs and clustering statistics at $z\sim6.6$
  (Ouchi et al.\ 2010: \textit{open blue square and circle}) and at
  $z\sim7$ (Ota et al.\ 2008: \textit{open blue circle}), and the
  prevalence of Ly$\alpha$ emission in $z\sim7$-8 galaxies (S14:
  \textit{large red circles}).  Also included here (\textit{small red
    solid circles}) are other estimated constraints on $Q_{HII}$ from
  the prevalence of Ly$\alpha$ emission from galaxies at $z\sim7$
  ($Q_{HII}<0.5$ [R13]; $Q_{HII}<0.4$ [Ono et al.\ 2012];
  $Q_{HII}<0.49$ [P14]; $Q_{HII}\sim0.5$ [Caruana et al.\ 2014]) and
  at $z\sim8$ ($Q_{HII}<0.7$ [Tilvi et al.\ 2014]).  The red and
  light-red-shaded region indicates the range of $Q_{HII}$ allowed for
  our models of the ionizing emissivity preferred at 68\% and 95\%
  confidence, respectively, and where reionization is complete between
  $z=5.9$ and $z=6.5$.  The magenta-hatched region indicates the range
  of $Q_{HII}$ allowed at 68\% confidence for the WMAP 9-year $\tau$
  measurement ($0.089\pm0.014$).\label{fig:reion}}
\end{figure*}

How does this luminosity density compare with a simple extrapolation
of the $z=4$-8 LF results to $z\sim10$?  Adopting the d$\log_{10}
\rho_{UV}/dz = -0.19\pm0.04$ scaling implied by the LF results of B15
(Table~\ref{tab:ldevol}), the extrapolated LF density at $z\sim10$ is
$10^{25.34\pm0.10}$ ergs s$^{-1}$ Hz$^{-1}$Mpc$^{-3}$ to $-17\,$mag
(see light-blue-shaded contour in Figure~\ref{fig:sfz}).  The
luminosity density we infer is consistent with this extrapolation
(similar to recent results by Coe et al.\ 2013, McLeod et al.\ 2015,
or the Oesch et al.\ 2015 results over the first Frontier Field).
However, it is also consistent at $1\sigma$ with the $(1+z)^{-10.8}$
evolution found by Oesch et al.\ (2014) at $z>8$, which suggests a
luminosity density $10^{25.1\pm0.3}$ ergs s$^{-1}$
Hz$^{-1}$Mpc$^{-3}$.  This is particularly the case, since $z>6$
galaxies may be more efficient at releasing ionizing radiation into
the IGM due to evolution in $f_{esc}$ or $\xi_{\rm ion}$ (both of
which have been speculated to increase at $z>6$: Siana et al.\ 2010,
Hayes et al.\ 2011, HM12, KF12, Duncan \& Conselice 2015).  This makes
the present estimate of the $UV$ luminosity density at $z\sim10$
effectively an upper limit.

\subsection{How the Ionizing Emissivities We Infer Compare with 
Key Observational Constraints}

Finally, it is useful to compare the results of our preferred models
for the evolution of the ionizing emissivity with the key
observational constraints we considered, as a check on the overall
self-consistency of the constraints.

The results for our models preferred at 68\% and 95\% confidence and
where reionization finishes at $z=5.9$-6.5 (i.e., when $Q_{HII}$
reaches 1 using Eq.~\ref{eq:m99}) are presented in
Figure~\ref{fig:reion}.  Both the preferred optical depths and
reionization histories are shown in this figure.  Our ionizing
emissivity models are in excellent agreement with all available
constraints, including those not considered in deriving the ionizing
emissivity (Table~\ref{tab:obsconst}).  A similar version of this
figure is presented in Appendix C, but also showing the constraints
based on the WMAP 9-year $\tau$'s.

As in other simple models of cosmic reionization (R13, R15), i.e.,
where inhomogeneities in the IGM are not treated, we are not able to
reproduce observations which suggest incomplete reionization to
$z=5$-6, i.e., $Q_{HII}\approx 1-10^{-4}$.  Correctly reproducing the
end of reionization would require careful simulations over
cosmologically-significant volumes, with voids and overdensities, and
require the treatment of radiative transfer effects.

Interestingly enough, the optical depths allowed at 68\% confidence
only include the lower 50\% of the values allowed by Planck and do not
significantly extend above $\tau$'s of 0.066.  This appears to be the
direct result of the constraint we apply on the filling factor of
ionized hydrogen at $z=8$ from S14.

If we do not consider the constraint from S14 and instead suppose that
$Q_{HII}=0.55\pm0.20$ at $z\sim8$ (resulting in an evolution of the
ionizing emissivity illustrated by the shaded-green region in the
lower-left panel of Figure~\ref{fig:back}), the Thomson optical depths
we derive span the range 0.060 to 0.082.  This demonstrates the value
of continued work on quantifying the prevalence of Ly$\alpha$ emission
in $z\sim8$ galaxies (e.g., S14) and also constraining the ionizing
state of the universe at $z\sim8$ using other methods (e.g., as Bolton
et al.\ 2011 do at $z\sim7.1$ using the damping wing of Ly$\alpha$
observed against the spectrum of a bright quasar).

\section{Discussion and Summary}

Here we have combined the new measurements of the Thompson optical
depth $\tau$ from Planck (PC15) with a new approach that focuses on
inferring the cosmic ionizing emissivity $\dot{N}_{\rm ion}$ and its
evolution over the reionization epoch from a variety of observables
(building of course on significant earlier work by e.g. Madau et
al.\ 1999; Bolton \& Haehnelt 2007; KF12; R13; Becker \& Bolton 2013;
R15: see also Mitra et al.\ 2011, 2012).  This approach has allowed us
to first gain insight into the allowed evolution of the sources
driving reionization, without immediately making assumptions about
their nature.  We then assess the implications of the derived
evolution for the ionizing emissivity and compare with the evolution
seen in the luminosity density $\rho_{UV}$ for galaxies.

We have derived empirical constraints on the evolution of the ionizing
emissivity based on the recently-measured Thomson optical depth $\tau$
(PC15) and the $Q_{HII}(z)$'s inferred from quasar absorption spectra
and the prevalence of Ly$\alpha$ emission in $z=7$-8 galaxies (Fan et
al.\ 2006a; McGreer et al.\ 2015; S14).  We tabulate in
Table~\ref{tab:obsconst} the extensive results on the filling factor
$Q_{HII}$ of ionized hydrogen from the literature that allow for us to
constrain the evolution of the ionizing emissivity $\dot{N}_{\rm ion}$
(see also Figure~\ref{fig:obs}).

We demonstrate that the evolution in the cosmic ionizing emissivity at
$z>6$, i.e., $\dndz=-0.15_{-0.11}^{+0.08}$
($\dndz=-0.19_{-0.11}^{+0.09}$ for a flat prior in $\dndz$ and
$\log_{10} \dot{N}_{\rm ion}(z=8)$), is matched by similar evolution
in the $UV$ luminosity density (after extrapolation to $-13$ mag),
i.e., d$\log_{10} \rho_{UV}/dz = -0.11\pm0.04$.  By contrast, use of
the 9-year WMAP optical depths $\tau=0.089\pm0.014$ (Bennett et
al.\ 2013) to derive the evolution in the emissivity yields
$\dndz=0.00_{-0.06}^{+0.03}$ (Table~\ref{tab:back}).

This is the first time this similar evolution has been shown
quantitatively in this manner and builds on the well-known case that
galaxies taken together can provide the UV ionizing radiation needed
for reionization (but see also R15).  This result further supports the
view that star formation in early galaxies drives the reionization of
the universe (\S4.2).

This conclusion is further strengthened by the fact that the requisite
conversion factor from the $UV$ luminosity density to the ionizing
emissivity ($\xi_{\rm ion} f_{esc} = 10^{24.50}$ s$^{-1}/($ergs
s$^{-1}$Hz$^{-1}$)) is consistent with physically plausible values for
the escape fraction $f_{esc}$ and $\xi_{\rm ion}$, for a faint-end
limit $M_{lim}$ to the $UV$ LF of $-13$ mag and a clumping factor
$C_{HII}=3$.  We calculate that this conversion factor has an
approximate uncertainty of $\sim$0.1 dex (\S4.2) based on the
uncertainties in the normalization of the cosmic ionizing emissivity
and the UV luminosity density.  We also present a generalization of
this constraint for other values of $M_{lim}$ and $C_{HII}$, i.e.,
Eq.~(\ref{eq:convf}).  We manipulate this constraint to provide a
general formula for deriving $f_{esc}$ for a wide range of different
values for $\xi_{ion}$, $M_{lim}$, $C_{HII}$ assuming that galaxies
drive the reionization of the universe (see Eq.~\ref{eq:fesc} and
Table~\ref{tab:fesc}).

This is also one of the first analyses where the uncertainty on this
conversion factor has been estimated from constraints on the
ionization state of the universe at $z>5.9$ and is the direct result
of the inferences we make regarding the evolution of the ionizing
emissivity (but see also Mitra et al.\ 2013).  [KF12 achieved a
  similar constraint on the conversion factor at $z\sim4$ from the
  ionizing emissivity estimates based on observations of the
  Ly$\alpha$ forest.]  Despite the size of these uncertainties, the
consistency with physically plausible values of $f_{esc}$, $\xi_{\rm
  ion}$, and $M_{lim}$ was not assured \textit{a priori} and provides
confidence that galaxies play a dominant role in reionization.

We also consider quasars/AGNs as potential sources of the ionizing
radiation.  However, as for most previous assessments, quasars/AGNs,
appear unlikely to be the dominant source of the ionizing UV
radiation, even under rather generous assumptions about the
contributions of faint AGNs.  There is simply little evidence they
show the required emissivities nor redshift dependence to match that
found for the ionizing emissivity (\S4.3: but see Giallongo et
al.\ 2015).

Assuming no change in the production efficiency of $UV$ continuum
photons or ionizing photons, our constraints on the evolution of the
ionizing emissivity can be used to estimate the $UV$ luminosity
density at $z\sim10$.  We show in \S4.4 that it is
$8_{-4}^{+15}\times$ lower than at $z\sim6$.  As we note, there is
also the possibility, due to either evolution in $\xi_{\rm ion}$ or
$f_{esc}$, that $z>6$ galaxies are more efficient (per unit $UV$
luminosity) at releasing Lyman-continuum radiation into the IGM than
$z\sim6$ galaxies.  If that is the case, then this estimate on the
$UV$ luminosity density of $z\sim10$ galaxies is an upper limit.  The
uncertainty on this estimate, however, is large enough that it cannot
help to resolve the question about the potential change in the slope
of the luminosity density at around $z\sim8$ as first identified by
Oesch et al.\ (2012).

The recent remarkable observations from Planck (PC15) provide a fresh
opportunity to re-evaluate the role of galaxies in cosmic reionization
(see also R15; Mitra et al.\ 2015).  We have taken advantage of these
new Planck results on the Thompson optical depth, as well as a decade
of observations of galaxies and quasars that provide constraints on
the filling factor of ionized hydrogen as a function of redshift (see
\S 4.5), to take a further step.  We show not only that reionization
is consistent with being driven by the UV radiation from galaxies, as
many others have demonstrated, but also that the evolution of the
ionizing emissivity (from $z\sim4$ to $z\sim10$) matches a similar
trend in the UV luminosity density (see Figure~\ref{fig:nnion1}).

The results here substantially strengthen the growing consensus that
early galaxies are the key to reionization.  These results on cosmic
reionization from current state-of-the-art microwave background probes
(PC15) combined with deep probes of faint $z>6$ galaxies provided by
the Hubble Space Telescope (e.g., Bouwens et al.\ 2011; McLure et
al.\ 2013; R13; B15; Atek et al.\ 2015) have demonstrated the power of
combining data from a wide range of major missions, and highlight the
upcoming opportunities with JWST for probing deeply into the
reionization epoch.

\acknowledgements

We are grateful to George Becker, Andrea Ferrara, Kristian Finlator,
Marijn Franx, Andrei Mesinger, Sourav Mitra, Joop Schaye, and Michele
Trenti for conversations related to this work.  This paper owes its
existence to Andrei Mesinger's tasking the first author with writing a
chapter for a future book.  We acknowledge the support of NASA grant
NAG5-7697, NASA/STScI grant HST-GO-11563, and a NWO vrij-competitie
grant 600.065.140.11N211.

\appendix

\section{A.  Production Efficiency for Lyman Continuum Photons Per Unit UV Luminosity}

In this section, we determine how the production efficiency per unit
$UV$ luminosity $\xi_{\rm ion}$ depends on the $UV$-continuum slope
$\beta$ (where $\beta$ is defined such that $f_{\lambda}\propto
\lambda^{\beta}$).  Given that observations only allow for a
constraint on the product of the two unknowns $\xi_{\rm ion}f_{esc}$
(and perhaps more generally the product of the four unknowns
$x_{ion} f_{esc} f(M_{lim}) (C_{HII}/3)^{-0.3}$: see
Eq.~\ref{eq:convf} from \S4.2), choice of a physically plausible value
for $\xi_{\rm ion}$ ensures that the $f_{esc}$ we infer is also of
more relevance.

\begin{figure}
\epsscale{0.7}
\plotone{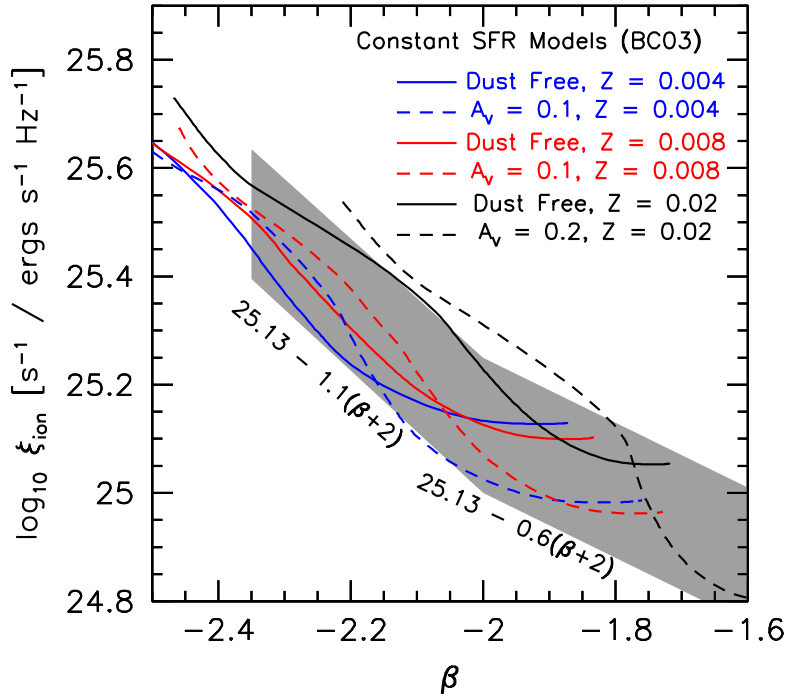}
\caption{A determination of how the production efficiency $\xi_{\rm
    ion}$ of Lyman-continuum photons per unit $UV$ luminosity at
  1600\AA$\,\,$depends on the $UV$-continuum slope $\beta$.  These
  efficiencies are calculated from the Bruzual \& Charlot (2003)
  spectral synthesis library assuming a constant star formation rate.
  We also adopt three different metallicities (0.2$Z_{\odot}$,
  0.4$Z_{\odot}$, and $Z_{\odot}$) and a wide range in ages (0.1 Myr
  to 10 Gyr).  Both the case of no dust content and $A_{V} = 0.1/0.2$
  (Charlot \& Fall 2000) is considered, as indicated on this figure.
  $\beta$ is computed over the spectral range 1700\AA$\,\,$to 2200\AA.
  The shaded envelopes indicate the approximate dependence of
  $\xi_{\rm ion}$ on $\beta$.  For completeness the full range of
  dusty and dust-free tracks are shown, even up to solar metallicity.
  We recognize that solar metallicities may be unlikely at $z>6$ where
  results (e.g., Bouwens et al.\ 2014; Dunlop et al.\ 2013) suggest
  that the dust content is very low.\label{fig:xi}}
\end{figure}

Following a similar procedure to that executed in R13, we can estimate
the efficiency parameter $\xi_{\rm ion}$ by considering a variety of
different ages, metallicities, and dust content for star-forming
galaxies at $z>6$.  For convenience, the star-formation rate of our
model galaxies is assumed to be constant.  The Charlot \& Fall (2000)
dust model is assumed, and we also leverage the spectral synthesis
models of Bruzual \& Charlot (2003).  The $UV$-continuum slope $\beta$
is derived from the model spectra over the wavelength 1700\AA$\,\,$and
2200\AA, consistent with the position of the broadbands used to derive
$\beta$ for $z\sim7$ galaxies (e.g., Bouwens et al.\ 2010, 2014;
Dunlop et al.\ 2013).  The conversion factors and $\beta$'s computed
for many different model spectra are presented in Figure~\ref{fig:xi}.

Fitting the envelope of derived conversion factors $\xi_{\rm ion}$ versus
$\beta$, we find $10^{25.13-1.1(\beta+2)}$ s$^{-1}/($ergs
s$^{-1}$Hz$^{-1}$) for $\beta<-2$ and $10^{25.13-0.6(\beta+2)}$
s$^{-1}/($ergs s$^{-1}$Hz$^{-1}$) for $\beta>-2$, with an approximate
width of this distribution of $\pm0.125$ in $\log_{10} \xi_{\rm ion}$.

\begin{figure*}
\epsscale{1.14}
\plotone{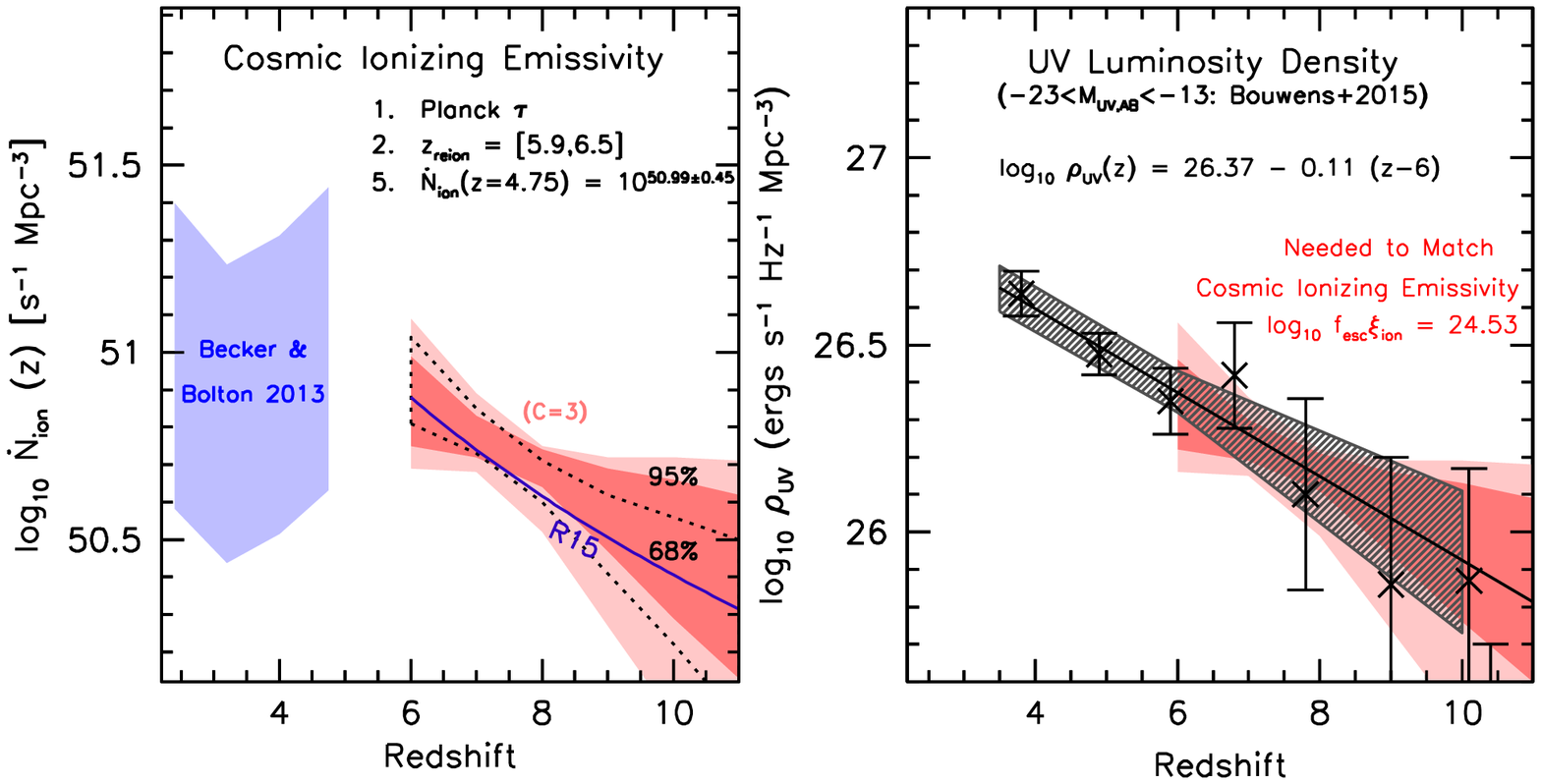}
\caption{(\textit{left}) 68\% and 95\% confidence intervals
  (\textit{red-shaded region}) on the evolution of the cosmic ionizing
  emissivity (assuming $C_{HII}=3$) supposing that reionization is
  complete at $z=5.9$-6.5, utilizing the latest Planck optical depth
  measurements (PL15), and enforcing continuity with inferred ionizing
  emissivity at $z=4.75$ (constraints 1, 2, and 5 from
  Table~\ref{tab:obsconst}) and excluding constraints based on the
  prevalence of Ly$\alpha$ emission in galaxies (Appendix B).  The
  dotted black lines demarcate the 68\% confidence regions for our
  fiducial determination.  See the caption to Figure~\ref{fig:nnion1}
  for a description of other symbols and regions.  (\textit{right})
  Comparison of the evolution of the inferred ionizing emissivity with
  the evolution of the $UV$ luminosity density (\textit{black-hatched
    region}) adopting some redshift-independent conversion factor
  $\log_{10}f_{esc}\xi_{\rm ion}=24.53$.  As before, the evolution of
  the ionizing emissivity is very similar to that found for the
  luminosity density of galaxies in the rest-frame $UV$.  While this
  again suggests that star-forming galaxies drive the reionization of
  the universe, this figure demonstrates that this conclusion does not
  depend on whether or not we make use of constraints that involve
  Ly$\alpha$ emission from $z>6$ galaxies.\label{fig:nnion1MI}}
\end{figure*}

\section{B.  Evolution Inferred for Cosmic Ionizing Emissivity Excluding Constraints Based on the Prevalence of Ly$\alpha$ Emission in $z>6$ Galaxies}

As discussed extensively in the text and elsewhere (e.g., Mesinger et
al.\ 2015; Choudhury et al.\ 2015), large uncertainties exist in the
use of Ly$\alpha$ emission from galaxies to constrain the ionization
state of the universe at $z>6$.  Not only do the observational
inferences depend on the intrinsic velocity offset of Ly$\alpha$
emission (e.g., Mesinger et al.\ 2015; Stark et al.\ 2015), but it is
dependent on details of the radiative transfer, inhomogeneities in the
IGM, the ionizing background, all of which rely on high-quality
simulations of the reionization process (and where it can be
challenging to include all of the relevant physics).

As such, it is perhaps useful to derive the evolution in the cosmic
ionizing emissivity without relying on observational constraints that
involve the prevalence of Ly$\alpha$ emission in $z\sim7$ and $z\sim8$
galaxies.  We therefore repeat the exercise we performed in \S4.1 but
excluding constraints 3-4 from Table~\ref{tab:obsconst}.

Our derived constraints on the evolution of the ionizing emissivity
(68\% and 95\% confidence intervals) are shown in the left panel of
Figure~\ref{fig:nnion1MI}.  The equivalent $\dndz$, after
marginalizing over parameter space and weighting by $|\nabla\tau|$ (a
flat prior in $\tau$: see \S4.1), is $-0.09_{-0.13}^{+0.06}$
(Table~\ref{tab:back}), which is 0.5$\sigma$ shallower than in our
fiducial analysis (\textit{dotted black line gives the 68\% confidence
  intervals}) where we fold in the constraints from the observed
Ly$\alpha$ fraction.  The right panel in Figure~\ref{fig:nnion1MI}
compares these emissivity results with the evolution of the $UV$
luminosity density (see also Figure~\ref{fig:nnion1}).

Here again the evolution we infer for the ionizing emissivity is in
excellent agreement with the evolution found in the $UV$ luminosity
density (B15).  As in our fiducial analysis, the similarity of the
evolution found for these two quantities suggests that star-forming
galaxies provide the ionizing photons which drive the reionization of
the universe.  The exercise we present in this appendix shows that
this conclusion does not depend on our use of constraints that involve
the prevalence of Ly$\alpha$ emission from $z>6$ galaxies.

\begin{figure*}
\epsscale{1.15} \plotone{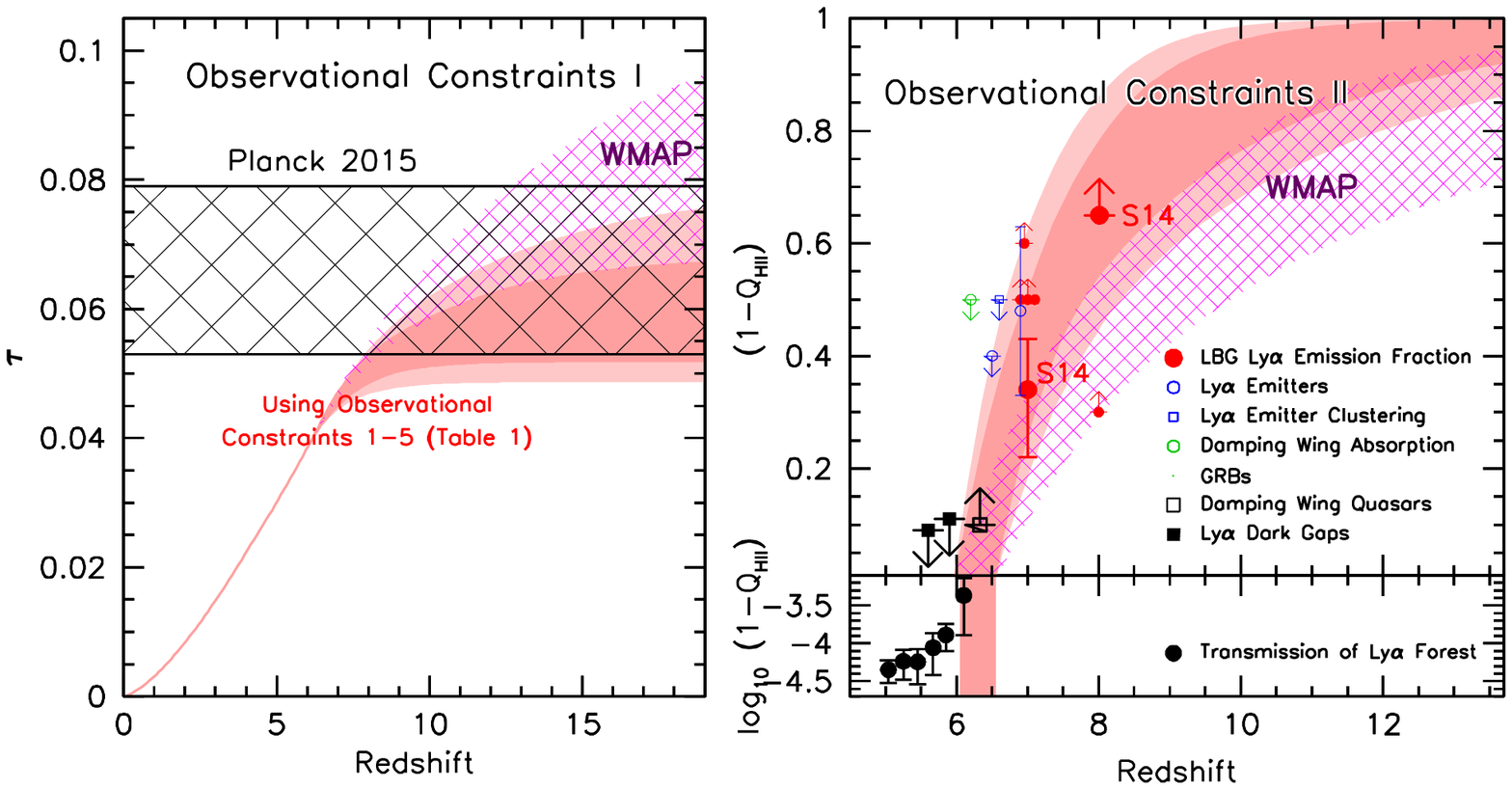}
\caption{(\textit{left}) Comparison of the constraints on the Thomson
  optical depth $\tau$ provided by the Planck 3-year results (PC15:
  \textit{cross-hatched black region}) and the cumulative optical
  depths preferred at 68\% and 95\% confidence by our models of the
  ionizing emissivity evolution including new constraints from Planck.
  The magenta-hatched region show the constraints that would be
  provided by the WMAP results (68\% confidence).  (\textit{right})
  Shown are constraints on the filling factor of ionized hydrogen
  $Q_{HII}$ as a function of redshift and the evolution preferred at
  68\% and 95\% confidence by our models of the ionizing emissivity
  evolution.  The magenta-hatched region indicates the range of
  $Q_{HII}$ allowed at 68\% confidence for the WMAP 9-year $\tau$
  measurement ($0.089\pm0.014$).  Similar to
  Figure~\ref{fig:reion}.\label{fig:reion2}}
\end{figure*}

\section{C.  Comparison between the Integrated $\tau$'s and Reionization Histories Implied by Planck and the WMAP 9-year Results}

For context and to illustrate the gains from Planck, we also
Figure~\ref{fig:reion2} showing the filling factor evolution of
ionized hydrogen, $Q_{HII}$, and integrated Thomson optical depths
$\tau$'s as inferred from the WMAP 9-year $\tau$ measurement and
assuming reionization is complete at $z=5.9$-6.5 (\textit{magenta
  hatched region}: 68\% confidence).

\end{document}